%% file: main.tex
\documentclass[acmsmall,nonacm]{acmart}

\AtBeginDocument{
  }

\makeatletter
\let\@authorsaddresses\@empty
\makeatother                                     
\renewcommand\footnotetextcopyrightpermission[1]{}

\usepackage{tikz}
\usepackage{braket}
\usepackage{lipsum}
\usepackage{subfig}
\usepackage{xcolor}
\usepackage{amsmath}
\usepackage{balance}
\usepackage{environ}
\usepackage{physics}
\usepackage{colortbl}
\usepackage{enumitem}
\usepackage{amsfonts}
\usepackage{booktabs}
\usepackage{graphicx}
\usepackage{textcomp}
\usepackage{algorithm}
\usepackage{algpseudocode}
\usepackage{arydshln}
\usepackage[normalem]{ulem}

\newcommand{\sol}{\textsc{Anchor}}

\newcommand{\rev}[1]{\textcolor{black}{#1}}

\begin{document}

\title{\sol{}: Reducing Temporal and Spatial Output Performance Variability on Quantum Computers}

\author{Yuqian Huo}\affiliation{\institution{Rice University}\country{Houston, TX, USA}}
\author{Daniel Leeds}\affiliation{\institution{Rice University}\country{Houston, TX, USA}}
\author{Jason Ludmir}\affiliation{\institution{Rice University}\country{Houston, TX, USA}}
\author{Nicholas S. DiBrita}\affiliation{\institution{Rice University}\country{Houston, TX, USA}}
\author{Tirthak Patel}\affiliation{\institution{Rice University}\country{Houston, TX, USA}}

\renewcommand{\shortauthors}{Yuqian Huo, Daniel Leeds, Jason Ludmir, Nicholas S. DiBrita, and Tirthak Patel}


\begin{abstract}

Quantum computing, which has the power to accelerate many computing applications, is currently a technology under development. As a result, the existing noisy intermediate-scale quantum (NISQ) computers suffer from different hardware noise effects, which cause errors in the output of quantum programs. These errors cause a high degree of variability in the performance (i.e., output fidelity) of quantum programs, which varies from one computer to another and from one day to another. Consequently, users are unable to get consistent results even when running the same program multiple times. Current solutions, while focusing on reducing the errors faced by quantum programs, do not address the variability challenge. To address this challenge, we propose \sol{}, a first-of-its-kind technique that leverages linear programming to reduce the performance variability by 73\% on average over the state-of-the-art implementation focused on error reduction.

\end{abstract}



\maketitle

\pagestyle{plain}

\input{sections/introduction}
\input{sections/background}
\input{sections/motivation}
\input{sections/design}
\input{sections/methodology}
\input{sections/evaluation}
\input{sections/related_work}
\input{sections/conclusion}

\balance

\bibliographystyle{ACM-Reference-Format}
\bibliography{main}

\appendix

\input{sections/appendix}

\end{document}

%% file: sections/introduction.tex
\section{Introduction}
\label{sec:introduction}

Quantum computing holds the potential to revolutionize a wide range of fields by solving certain classes of problems exponentially faster than classical computers~\cite{preskill2018quantum,preskill2021quantum,bova2021commercial}. This emerging technology leverages the principles of quantum mechanics, such as superposition and entanglement, to perform computations in fundamentally new ways. Quantum computing cloud services like AWS Braket~\cite{aws}, IBM Quantum~\cite{castelvecchi2017ibm}, and Azure Quantum~\cite{hooyberghs2022azure} have expanded access to quantum hardware, enabling researchers and developers to run quantum algorithms on real quantum processors. Despite its promise, quantum computing is still in the developmental stage, with current and near-future devices being characterized by their susceptibility to various forms of hardware noise, which introduces errors in quantum computations~\cite{farhi2016quantum,ramezani2020machine,liu2021rigorous, jerbi2023quantum,liu2024towards}.

Superconducting qubits are the leading technology currently, leveraging superconductivity to create discrete energy levels that can be manipulated to represent qubit states. However, these qubits suffer from various noise effects, including decoherence and gate errors, which degrade the performance of quantum circuits. Manufacturing defects and variability in fabrication processes further exacerbate these issues, leading to inconsistencies in qubit parameters such as frequencies and anharmonicities~\cite{Krantz2019,McKay2017}. As a result, the output fidelity of quantum programs can vary significantly, both temporally (over different days on the same machine) and spatially (across different machines on the same day). 

This noise variability especially adversely affects iterative and variational quantum algorithms such as the quantum approximate operator ansatz (QAOA), the variational quantum eigensolver (VQE), and quantum machine learning (QML) algorithms. These algorithms require the circuits that are run in each iteration to provide an accurate estimate of the objective function, loss value, and gradient signal relative to the circuits run during other iterations. Noise variability severely limits this estimation capability, as the circuit runs during different iterations could be differently impacted by noise. Thus, the variability in hardware noise effects poses a significant challenge for quantum cloud services. Users often experience substantial fluctuations in the performance of their quantum circuits, which can lead to inconsistent and unreliable results~\cite{ravi2021quantum,li2023qasmbench,patel2020experimental,tomesh2022supermarq}.

Current techniques aimed at mitigating errors in quantum circuits primarily focus on improving the output fidelity between the ideal and actual output distributions, based on noise information available at the time of execution~\cite{tannu2019ensemble,tannu2019not,wille2019mapping,zulehner2019compiling,patel2020ureqa,patel2020veritas,gokhale2020optimized,li2019tackling}. While these approaches can reduce the average error, they do not adequately address the issue of variability in output fidelity caused by temporal and spatial factors. Temporal variability occurs when the performance of a quantum circuit varies substantially over different days due to dynamic changes in the noise environment, such as temperature fluctuations, crosstalk, and hardware recalibration cycles. Spatial variability arises from differences in noise profiles and qubit characteristics across different quantum computers within the same cloud service~\cite{kjaergaard2019superconducting,Krantz2019,McKay2017}.

\textit{\textbf{To address these challenges, we propose \sol{}\footnote{\sol{} is published in the Proceedings of the ACM SIGMETRICS International Conference on Measurement and Modeling of Computer Systems (SIGMETRICS), 2026.}, a novel technique designed to minimize the variability in the output fidelity of quantum circuits.}} To achieve this, \sol{} departs from the conventional practice of running all circuit experiments (or shots) on the same circuit map and introduces the idea of dividing the circuit shots across multiple different maps. It designs a linear programming framework to efficiently optimize this shot assignment, taking into account stochastic noise effects across both temporal and spatial dimensions. \sol{} also develops a learning-based predictor to predict output fidelity values as input for the linear programming framework. By doing so, \sol{} aims to provide more consistent and reliable results for users, regardless of the specific quantum hardware used or the time of execution.

\vspace{2mm}

\noindent\textbf{The key contributions of this work are as follows:} 
\begin{itemize}[leftmargin=*]
    \item We identify and quantify the impact of temporal and spatial variability on the performance of quantum circuits in the IBM Quantum cloud-based service.
    \item We propose \sol{}, which breaks away from the idea of optimizing only for the output fidelity of the circuit at any given time on any given computer and focuses on optimizing the variability of the output fidelity across multiple times and computers. 
    \item We formulate the implementation of \sol{} in an efficient linear programming representation and develop a low-overhead learning-based predictor to generate output error. Our implementation framework and experimental data will be open-sourced for community-wide adoption and future research development. 
    \item We demonstrate the effectiveness of \sol{} through simulation and real-computer experiments, showing a significant reduction of 73\% in performance variability.
    \item We also provide a comprehensive overhead and ablation analysis to demonstrate the efficacy of \sol{}'s implementation to enhance the reliability of quantum cloud environments.
    \item \rev{\sol{}'s framework, codebase, and experimental data are open-sourced for reproducibility and research community adoption: \textit{\url{https://github.com/positivetechnologylab/Anchor}}.}
\end{itemize}

\vspace{2mm}

\noindent\textbf{Paper Organization:} The remainder of this paper is organized as follows: Sec.~\ref{sec:background} provides the relevant background of quantum computing. Sec.~\ref{sec:motivation} motivates the need for \sol{} by presenting variability results and Sec.~\ref{sec:design} presents the design of \sol{}. Then, Sec.~\ref{sec:methodology} provides the experimental methodology, followed by Sec.~\ref{sec:evaluation} presenting the evaluation results and analysis. Finally, Sec.~\ref{sec:related_work} discusses the related work, and Sec.~\ref{sec:conclusion} concludes the paper and discusses its potential impact.

%% file: sections/background.tex
\section{Background}
\label{sec:background}

This section presents a brief, relevant background of quantum computing concepts and technology.

\vspace{2mm}

\noindent\textbf{The Fundamentals of Quantum Computing:} A \textit{qubit} is the fundamental unit of quantum information, analogous to a classical bit but with quantum mechanical properties. Unlike a classical bit, which can be in one of two definite states, 0 or 1, a qubit can simultaneously exist in a \textit{superposition} of both states. Mathematically, the state of a qubit is represented as $\ket{\psi} = \alpha\ket{0} + \beta\ket{1}$, where $\alpha$ and $\beta$ are complex numbers satisfying $|\alpha|^2 + |\beta|^2 = 1$. When dealing with multiple qubits, the state of the system is described by a tensor product of the individual qubit states. For instance, a two-qubit state is given by $\ket{\psi} = \alpha_{00}\ket{00} + \alpha_{01}\ket{01}+ \alpha_{10}\ket{10} + \alpha_{11}\ket{11}$, where $\alpha_{ij}$ are complex coefficients satisfying the normalization condition $\sum_{i,j} |\alpha_{ij}|^2 = 1$. Each $|\alpha_{ij}|^2$ gives the probability of measuring the corresponding state $ij$. Quantum gates perform operations on qubits analogous to classical gates on bits. The U3 gate is a one-qubit gate that helps create the desired superposition using three parameterized angles: $\theta$, $\phi$, and $\lambda$. The U3 gate can be used to generate commonly used one-qubit gates such as X, SX, and Rz. For two-qubit operations that help \textit{entangle} qubits, the controlled-X (CX or CNOT) gate is used. It acts on a pair of qubits, flipping the state of the target qubit if the control qubit is in the $\ket{1}$ state. The two gates have matrix representations as follows.

\[
U3(\theta, \phi, \lambda) = \begin{pmatrix}
\cos(\theta/2) & -e^{i\lambda}\sin(\theta/2) \\
e^{i\phi}\sin(\theta/2) & e^{i(\phi + \lambda)}\cos(\theta/2)
\end{pmatrix} \ \ \ \
CX = \begin{pmatrix}
1 & 0 & 0 & 0 \\
0 & 1 & 0 & 0 \\
0 & 0 & 0 & 1 \\
0 & 0 & 1 & 0
\end{pmatrix}
\]

\begin{figure}
    \centering
    \subfloat[Quantum Circuit]{\includegraphics[scale=0.41]{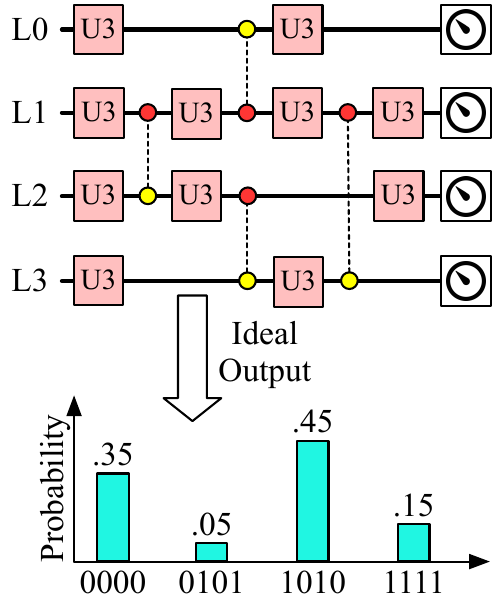}}
    \hfill
    \subfloat[Computer Layout]{\includegraphics[scale=0.41]{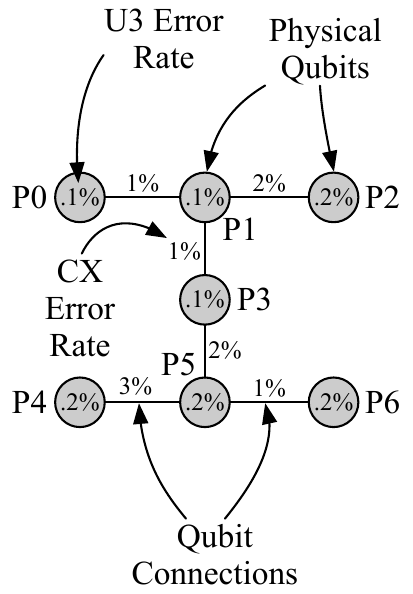}}
    \hfill
    \subfloat[Circuit Map A]{\includegraphics[scale=0.41]{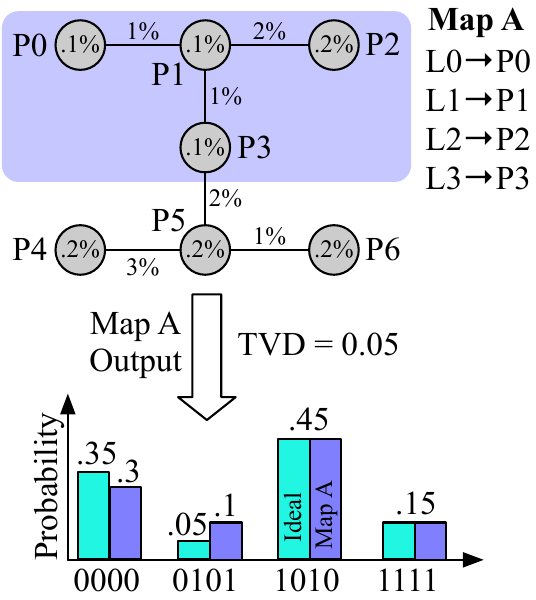}}
    \hfill
    \subfloat[Circuit Map B]{\includegraphics[scale=0.41]{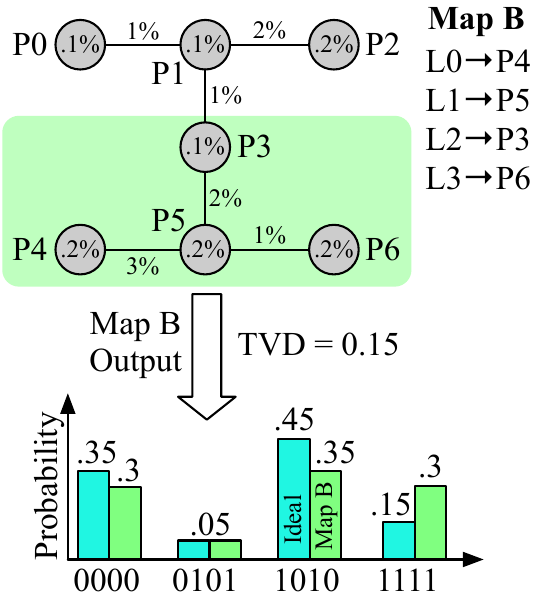}}
    \vspace{1mm}
    \hrule
    \vspace{-3.5mm}
    \caption{(a) An example four-qubit quantum circuit. L0-L3 are the four ``logical'' qubits to which one-qubit U3 and two-qubit CX gates are applied from left to right. For CX gates, the red dot indicates the control qubit, and the yellow dot indicates the target qubit. At the end of the computation, measurement gates are applied, which then generate one output state (e.g., $\ket{1010}$). An output probability distribution over all the states is generated when the circuit is run and measured multiple times (e.g., 1024 times or ``shots''). This circuit's ideal output distribution is shown when run on a noise-free quantum computer. (b) The layout of an example quantum computer with seven ``physical'' qubits with different error rates due to hardware noise. (c)-(d) Impact of the hardware noise effects when the circuit is run on two different circuit maps: Map A with lower error rates causes less output error (TVD), and Map B with higher error rates causes more TVD.}
    \label{fig:circuit}
    \vspace{-4mm}
\end{figure}

These quantum gates form a universal basis and are combined to form quantum circuits (shown in Fig.~\ref{fig:circuit}(a)), which are then used to perform computations by exploiting the principles of quantum superposition and entanglement. Once an $n$-qubit quantum circuit is executed, the qubits are measured, at which point the qubit superposition \textit{collapses} and the measurement manifests as one of the $2^n$ states. \rev{A \textit{shot} refers to a single execution of the entire quantum circuit followed by measurement of all qubits. Because each shot yields only one outcome from the underlying probability distribution, multiple \textit{shots} are typically run to empirically estimate the output probability distribution and enable algorithmic purposes.}

\vspace{2mm}

\noindent\textbf{Superconducting Qubit Technology:} Superconducting qubit technology is one of the leading platforms for realizing quantum computation due to its scalability and coherence properties. Superconducting qubits, also known as Josephson junction qubits, leverage the principles of superconductivity to create discrete energy levels that can be manipulated to represent qubit states. These qubits are fabricated using conventional lithographic techniques, allowing integration into complex circuits. The state of a superconducting qubit is controlled using microwave pulses, and longer qubit coherence times can be achieved through advanced materials and designs, such as transmon qubits~\cite{Koch2007,Barends2014}. Furthermore, superconducting qubits can be coupled via microwave resonators, enabling entanglement and the implementation of two-qubit gates~\cite{Blais2004}. Recent advancements in error mitigation and quantum algorithms on superconducting qubit platforms demonstrate the potential of this technology for the NISQ era of quantum computing~\cite{arute2019quantum}.

\vspace{2mm}

\noindent\textbf{Hardware Noise Effects on Quantum Computers:} Superconducting quantum hardware is susceptible to various noise effects that can significantly impact the performance and reliability of quantum computations~\cite{kjaergaard2019superconducting}. Manufacturing defects and variability in the fabrication process can introduce imperfections in superconducting qubits and their couplings~\cite{Krantz2019,McKay2017}. These imperfections manifest as variations in qubit frequencies, anharmonicities, and coupling strengths, which can lead to inhomogeneous broadening and decoherence. Even slight deviations in the fabrication process can cause significant discrepancies in the qubit parameters, affecting their coherence times and gate fidelities. As shown in Fig.~\ref{fig:circuit}(b), while one-qubit gates like the U3 gate typically exhibit low error rates due to their shorter operation times and lower complexity, two-qubit gates such as the CX (CNOT) gate suffer from higher error rates~\cite{das2019case,Baker2021-wv,patel2020experimental}. This is because two-qubit gates involve entangling operations that are more sensitive to noise and crosstalk, exacerbating the effects of fabrication variability and environmental disturbances~\cite{kjaergaard2019superconducting,Krantz2019,McKay2017}.

The presence of noise and errors in superconducting quantum circuits leads to deviations in the output distribution from the ideal, error-free case. To quantify the impact of these errors, the total variation distance (TVD) metric is commonly used~\cite{patel2021qraft,patel2022charter}. TVD measures the difference between the probability distributions of the ideal output and the experimentally observed erroneous output. Mathematically, TVD is defined as $\text{TVD}(P, Q) = \frac{1}{2} \sum_{x} |P(x) - Q(x)|\times{}100\%$, where $P(x)$ and $Q(x)$ are the probabilities of observing outcome $x$ in the ideal and erroneous distributions, respectively. A higher TVD (e.g., 50\%) indicates a greater deviation from the ideal distribution, signifying a lower fidelity of the quantum circuit. By analyzing the TVD, researchers can assess the fidelity of quantum operations and identify sources of noise and error in superconducting qubit systems, aiding in the development of error mitigation strategies~\cite{Wallman2015,Gambetta2012}.

\vspace{2mm}

\noindent\textbf{Quantum Circuit Mapping for Error Mitigation:} Connecting the understanding of noise effects on superconducting qubits to practical applications, significant efforts have been directed toward improving quantum circuit performance on NISQ computers. A key focus has been on optimal mapping and routing of quantum circuits to minimize output errors based on stochastically-varying qubit error rates measured during calibration cycles (typically, once or twice per day)~\cite{tannu2019ensemble,tannu2019not,wille2019mapping,zulehner2019compiling,patel2020ureqa,patel2020veritas,gokhale2020optimized,li2019tackling}. One pioneering work by Tannu et al.~\cite{tannu2019not} emphasized the variation in error rates among qubits, advocating for strategic mapping to minimize overall circuit error. This approach involves selecting qubits with lower error rates and placing critical gates on these qubits to enhance the circuit's fidelity. Additionally, routing involves determining the optimal paths for two-qubit gates to mitigate crosstalk interference and other noise sources. As demonstrated in the examples shown in Fig.~\ref{fig:circuit}(c) and Fig.~\ref{fig:circuit}(d), choosing a circuit map with lower error rates to run a given quantum circuit helps achieve an overall lower TVD in the output, improving the output fidelity. Thus, it is common for cloud services such as IBM Quantum to apply state-of-the-art circuit mapping techniques before running the circuits on their quantum computers~\cite{aleksandrowiczqiskit}.

\vspace{2mm}

\begin{figure}
    \centering
    \includegraphics[scale=0.38]{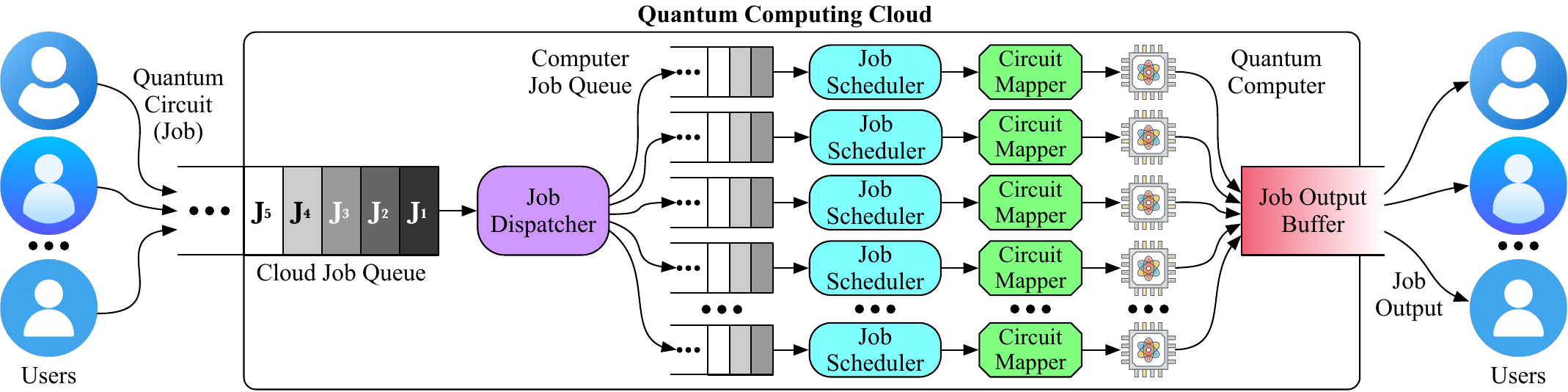}
    \vspace{1mm}
    \hrule
    \vspace{-3.5mm}
    \caption{When a user submits a job to the quantum computing cloud, the job is first added to a queue and dispatched to one of the computers in the cloud. Each computer has its own queue of jobs, from which jobs are scheduled and mapped to a region within the computer. After running the desired number of shots, the output probability distribution of the job is returned to the user.}
    \label{fig:cloud}
    \vspace{-4mm}
\end{figure}

\noindent\textbf{Quantum Computing Cloud Services:} Quantum computers are now offered in the form of cloud services: Quantum as a Service (QaaS). These include cloud resources like AWS Braket~\cite{aws}, IBM Quantum~\cite{castelvecchi2017ibm}, and Azure Quantum~\cite{hooyberghs2022azure}. Fig.~\ref{fig:cloud} shows the workflow of quantum cloud services. The structure is similar to classical cloud services, except for the additional step of optimally mapping the circuit once a job is scheduled on a quantum computer to reduce the impact of hardware noise effects. This work assumes the commonly used first-come, first-served (FCFS) policy for the job dispatcher and the job scheduler~\cite{li2021ribbon,grosof2023new,zhang2024cross}.

Any job that arrives is dispatched to the quantum computer with the lowest current job load, and the next job that a given computer schedules is the job that arrived first. Thus, the first job that arrives has the earliest chance of getting executed. Some quantum cloud computing services offer the user the ability to schedule their jobs on specific computers; however, this is untenable going forward as it results in hours or even days of queue times on the computers perceived as having the lowest errors~\cite{ravi2021quantum,nguyen2024quantum,zhang2022research}. Thus, quantum cloud service providers are likely to move to the classical FCFS model. Nonetheless, the exact job dispatching and job scheduling policy is not of relevance to our work and does not affect the quality of our results.

%% file: sections/motivation.tex
\section{Motivation for \sol{}}
\label{sec:motivation}

\begin{figure}
    \centering
    \subfloat[Temporal Variability]{\includegraphics[scale=0.47]{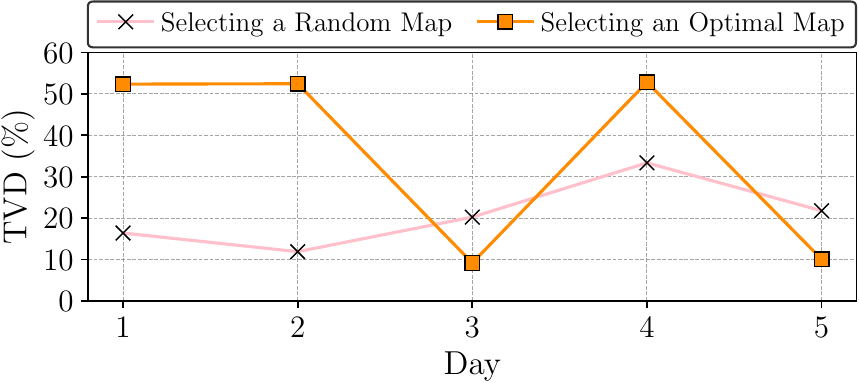}}
    \hfill
    \subfloat[Spatial Variability]{\includegraphics[scale=0.47]{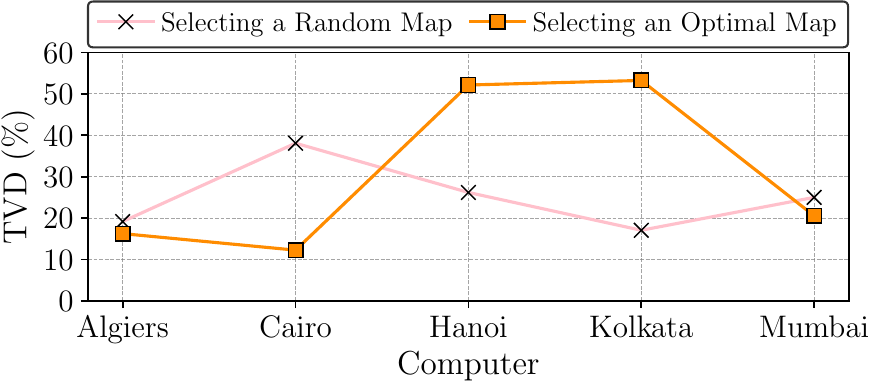}}
    \vspace{1mm}
    \hrule
    \vspace{-3.5mm}
    \caption{(a) Temporal and (b) spatial variability in the TVD of a quantum circuit when the 4-qubit VAR algorithm is run on the IBM quantum computers. We use the IBM QASM simulator to generate the ideal and noisy results. We use the same routing algorithms (routing determines how the circuit is laid out on the selected qubit map) for both map selection techniques. Refer to Sec.~\ref{sec:methodology} for the complete methodology.}
    \label{fig:motiv}
    \vspace{-4mm}
\end{figure}

In this section, we discuss the performance variability challenge in quantum computing as a result of the noise effects faced by the quantum hardware. Quantum computing cloud services have enabled broader access to quantum hardware, facilitating advancements in quantum algorithm development and experimentation. However, the variability in the performance of quantum circuits due to noise effects poses a significant challenge. Current approaches primarily focus on minimizing the TVD based on available noise information at the time of execution, but they do not address the issue of variability in output fidelity. This variability can arise both temporally (when running the same circuit on the same computer over multiple days) and spatially (when running the same circuit on different computers on the same day). Understanding and mitigating this variability is crucial for ensuring reliable quantum computations and enhancing the user experience.

\vspace{2mm}

\noindent\textbf{Temporal Variability in Performance on Different Days:} Temporal variability refers to the fluctuations in the performance of a quantum circuit when executed on the same quantum computer (e.g., $ibm\_hanoi$) over different periods. Fig.~\ref{fig:motiv}(a) illustrates this variability, showing how the output fidelity of a quantum circuit can change significantly over multiple days. The figure shows the TVD when the optimal circuit map is selected for the VAR algorithm on the $ibm\_hanoi$ computer by applying Qiskit's state-of-the-art noise-adaptive optimizations to a circuit on a given quantum computer based on the noise information available from the current calibration cycle~\cite{aleksandrowiczqiskit}. The figure also shows the TVD when any random circuit map is generated to run the circuit. As the figure shows, both techniques have high variability, and in fact, the optimal mapping technique has higher variability in the TVD compared to the random map selection technique due to its reliance on unreliable noise instrumentation, which can sometimes produce worse than random results.

This stochastic TVD fluctuation is primarily due to the dynamic nature of the noise environment in superconducting qubits, influenced by factors such as temperature fluctuations, crosstalk, and hardware calibration cycles~\cite{ravi2021quantum,li2023qasmbench,patel2020experimental,tomesh2022supermarq}. While current techniques aim to reduce the TVD based on noise information available at the time of execution, they do not account for these temporal changes in noise characteristics. Consequently, the fidelity of the circuit's output can vary substantially from day to day, leading to inconsistent results for users. \textit{\textbf{A technique that can minimize this temporal variability would ensure more stable and reliable outputs over time, enhancing the robustness of quantum cloud services.}}

\vspace{2mm}

\noindent\textbf{Spatial Variability in Performance on Different Quantum Computers:} Spatial variability, on the other hand, refers to the differences in output fidelity when the same quantum circuit is run on different quantum computers within the same cloud service. Fig.~\ref{fig:motiv}(b) depicts this spatial variability, showing how the performance of a circuit can vary across different hardware platforms on the same day. This variability arises because different quantum computers can have distinct noise profiles, qubit quality, and coupling strengths, even when they belong to the same provider~\cite{patel2020ureqa,li2023qasmbench,patel2020experimental,tomesh2022supermarq}. Current methodologies that focus solely on minimizing TVD based on instantaneous noise information fail to address the discrepancies between different machines. This can be a significant challenge, especially when cloud service providers start disallowing users from choosing specific computers to load balance and reduce congestion, leading to unpredictable performance variations. \textit{\textbf{A technique that reduces spatial variability would enable cloud providers to deliver more uniform and dependable computational results, regardless of the specific quantum hardware used, thereby improving the service of quantum clouds.}}

To provide a more consistent and reliable quantum computing experience, \sol{} aims to address the challenges associated with both temporal and spatial variability. \sol{} would enable cloud providers to execute any quantum circuit on any available computer while maintaining low variability in the output fidelity, ultimately benefiting users with more predictable and stable performance. Next, we describe the design of \sol{}.

%% file: sections/design.tex
\section{\sol{}'s Design}
\label{sec:design}

We first provide an overview of the design decisions involved in the implementation of \sol{}.

\vspace{2mm}

\begin{figure}
    \centering
    \includegraphics[scale=0.36]{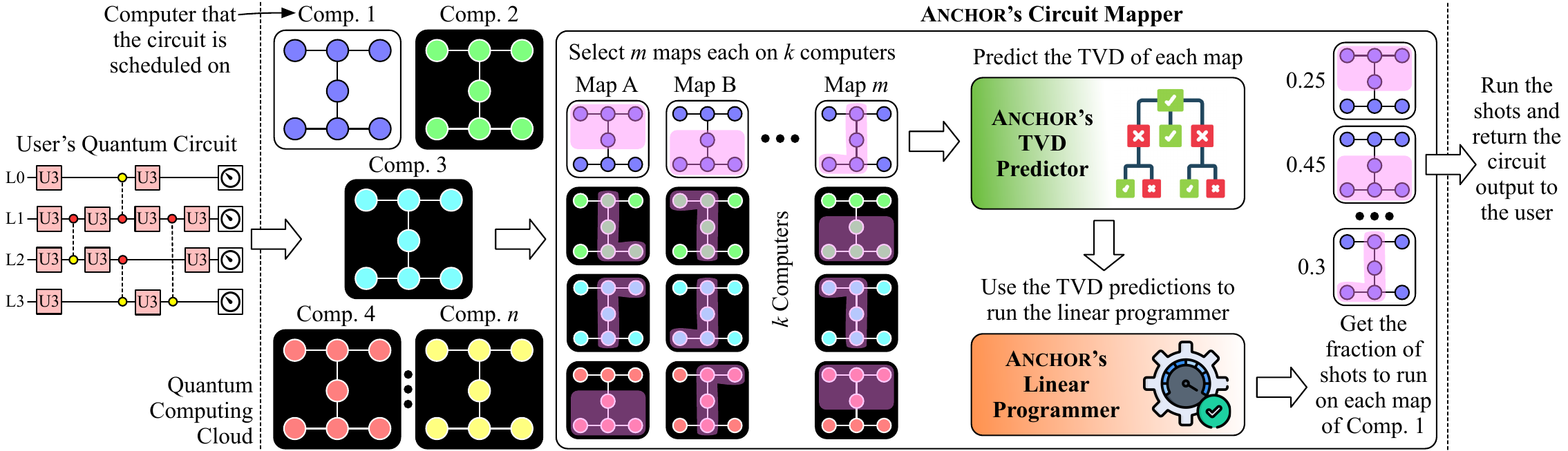}
    \vspace{1mm}
    \hrule
    \vspace{-3.5mm}
    \caption{Overview of the design of \sol{}'s quantum circuit mapper for reducing performance variability.}
    \label{fig:anchor}
    \vspace{-4mm}
\end{figure}

\noindent\textbf{Overview of the \sol{} Design:} \noindent\sol{} consists of three major components: the Linear programmer, the TVD Predictor, and the Circuit Maps Generator. The entire \sol{} pipeline is illustrated in Fig~\ref{fig:anchor}. Consider a scenario where a user has a quantum circuit $QC$. There are $k$ possible computers $Comp_{1}, Comp_{2}, \ldots, Comp_{k}$ with different physical qubit characteristics on which this $QC$ can be executed. If $Comp_{1}$ has the least waiting time or the user opts to run their $QC$ on it, the conventional method would involve transpiling (mapping and routing) the circuit on $Comp_{1}$ using the state-of-the-art circuit map techniques to generate a single map provided by the cloud provider. 

\sol{} proposes distributing the shots across different maps. Specifically, we have designed a linear programmer to approximately achieve the same TVD across various computers (but eventually running the circuit on only $Comp_{1}$) by considering the expected TVD of circuit maps on each computer while also minimizing the overall TVD of the circuit. \sol{} employs the following approach: (1) Circuit Maps Generator: Instead of allocating all shots to a single map, which is vulnerable to temporal and spatial influences, we generate $m$ different maps and distribute shots across them. (2) TVD Predictor: For each of the $m$ maps, \sol{}'s TVD Predictor estimates the TVDs for all maps across the $k$ computers and is trained to reduce temporal variability. (3) Linear Programmer: Using the predicted TVDs, the linear programmer calculates a shot distribution across different maps, such that the TVD for the shot distribution on $Comp_{1}$ is approximately equal to the TVD on the other computers, reducing spatial variability. This enables \sol{} to achieve the same TVD performance regardless of the chosen computer and time of run. After determining the shot distribution, the shots are run on the corresponding maps, and the total shot distributions are collected to obtain the final TVD. Next, we delve into the details of these design decisions.

\vspace{2mm}

\begin{figure}
    \centering
    \includegraphics[scale=0.365]{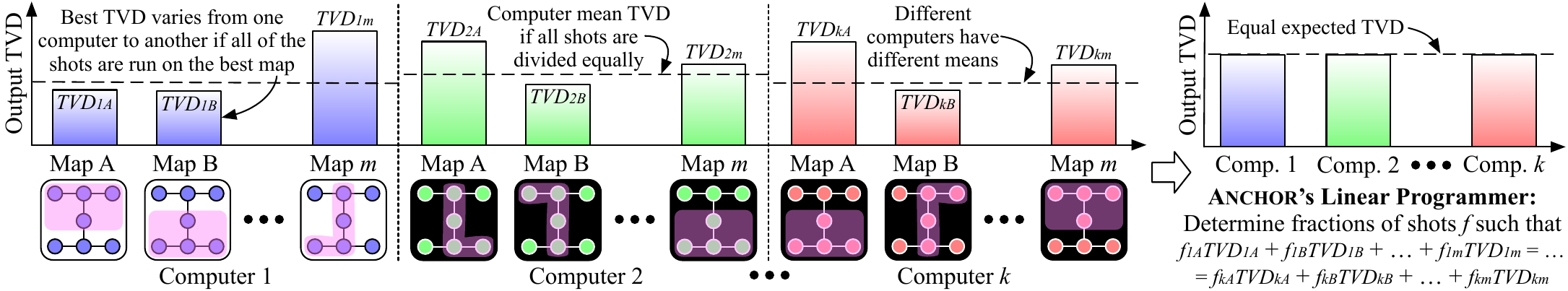}
    \vspace{1mm}
    \hrule
    \vspace{-3.5mm}
    \caption{\sol{}'s linear programmer divides shots across maps in a way that equalizes TVD across computers.}
    \label{fig:linear_prog}
    \vspace{-4mm}
\end{figure}

\noindent\textbf{The \sol{} Linear Programmer.} Distributing all shots onto the best map, as shown previously in Fig~\ref{fig:motiv}, results in significant TVD variations both temporally and spatially. An alternative, simple approach is to divide shots equally across different maps, but this also fails to address this issue effectively, as different computers still have varying mean TVDs (we demonstrate this in Sec.~\ref {sec:evaluation}). Thus, \sol{} needs to optimize the manner in which shots are distributed to different maps. More specifically, \sol{} formulates this optimization problem as a linear programming problem because linear programming is known to be computationally efficient at finding the optimal solutions~\cite{vanderbei1998linear}. Thus, leveraging a linear programming formulation allows \sol{} to make shot distribution decisions on the fly within seconds for each circuit as it is submitted to the cloud.

The rationale of \sol{} for reducing variance in the TVD of the output probability distribution for a given $QC$ is illustrated in Fig~\ref{fig:linear_prog}. Let $f_{ij}$ represent the fraction of shots for $Comp_{i}$ with map $j$, and let $\mathrm{TVD}_{ij}$ denote the TVD of $QC$ running on $Comp_{i}$ with map $j$. \sol{} proposes a method for calculating $f_{ij}$s to achieve an approximately equal output TVD across all $k$ computers. To achieve this, we establish two primary restrictions for the linear programmer:
\begin{itemize}
    \item The sum of shot distribution fractions for each computer should equal 1.
    \item The sum of $f_{ij} \cdot \mathrm{TVD}_{ij}$ for each computer should be approx. the same, as shown in Eq.~\ref{eq:linear_programming}.
\end{itemize}
{
\begin{equation}
\begin{aligned}
& f_{1A} \cdot \mathrm{TVD}_{1A} + f_{1B} \cdot \mathrm{TVD}_{1B} + \cdots + f_{1m} \cdot \mathrm{TVD}_{1m} \\
\approx & f_{2A} \cdot \mathrm{TVD}_{2A} + f_{2B} \cdot \mathrm{TVD}_{2B} +  \cdots + f_{2m} \cdot \mathrm{TVD}_{2m} \\
\vdots \\
\approx & f_{kA} \cdot \mathrm{TVD}_{kA} + f_{kB} \cdot \mathrm{TVD}_{kB} + \cdots + f_{km} \cdot \mathrm{TVD}_{km}
\end{aligned}
\label{eq:linear_programming}
\end{equation}}

However, solving for these constraints would only ensure that the overall TVD on each computer would be equal, but not that the TVD would be minimized as much as possible. \textit{The key idea behind the \sol{} linear programmer is to balance the predicted TVD values across different computers while minimizing the overall TVD.} \sol{}'s linear programmer uses the following formulation: 
\[
\text{minimize} \quad \mathbf{c}^T \mathbf{x}
\]

\noindent where {\small \(\mathbf{c}^T = [\mathrm{TVD}_{1A}, \mathrm{TVD}_{1B}, \ldots, \mathrm{TVD}_{1m}, \mathrm{TVD}_{2A}, \mathrm{TVD}_{2B}, \ldots, \mathrm{TVD}_{2m}, \ldots, \mathrm{TVD}_{kA}, \mathrm{TVD}_{kB}, \ldots, \mathrm{TVD}_{km}]\)}.

This minimization is subject to the constraints $\mathbf{A}_{eq} \mathbf{x} = \mathbf{b}_{eq}$ and $\mathbf{A}_{ub} \mathbf{x} \leq \mathbf{b}_{ub}$. Here, \(\mathbf{c}\) is the objective function coefficient vector that includes the TVD values of each circuit map. The equality constraints \(\mathbf{A}_{eq} \mathbf{x} = \mathbf{b}_{eq}\) ensure that the sum of shot distributions for each computer equals 1. The inequality constraints \(\mathbf{A}_{ub} \mathbf{x} \leq \mathbf{b}_{ub}\) ensure that the sum of \(\mathrm{TVD}_{ij}\) multiplied by the corresponding fraction satisfies the defined inequalities, and the bounds parameter defines the limits for each variable. Combining the two constraints, we get the following compound coefficient matrix \(\mathbf{A}\):

{\scriptsize
\[
\mathbf{A} = \begin{bmatrix}
1 & \dots & 1 & 0 & 0 & \dots & 0 & 0 \\
0 & 0 & \dots & 1 & 1 & \dots & 0 & 0 \\
\vdots & & & & \vdots & & & \vdots \\
0 & 0 & 0 & \dots & \dots & 1 & 1 & 1\\ 
\arrayrulecolor{blue}\hdashline
\mathrm{TVD}_{1A} & \mathrm{TVD}_{1B} & \dots & -(1+\epsilon)\mathrm{TVD}_{2A} & -(1+\epsilon)\mathrm{TVD}_{2B} & & \dots & 0 \\
-\mathrm{TVD}_{1A} & -\mathrm{TVD}_{1B} & \dots & \frac{1}{(1+\epsilon)}\mathrm{TVD}_{2A} & \frac{1}{(1+\epsilon)}\mathrm{TVD}_{2B} & & \dots & 0 \\
\vdots & & & & \vdots & & & \vdots \\
0 & 0 & \dots & \mathrm{TVD}_{(k-1)A} & \mathrm{TVD}_{(k-1)B} & \dots & -(1+\epsilon)\mathrm{TVD}_{k(m-2)} & -(1+\epsilon)\mathrm{TVD}_{k(m-1)} \\
0 & 0 & \dots & -\mathrm{TVD}_{(k-1)A} & -\mathrm{TVD}_{(k-1)B} & \dots & \frac{1}{(1+\epsilon)}\mathrm{TVD}_{k(m-2)} & \frac{1}{(1+\epsilon)}\mathrm{TVD}_{k(m-1)}
\end{bmatrix}
\]
}
{\scriptsize
\[
\mathbf{x} = \begin{bmatrix}
f_{1A} \\
f_{1B} \\
f_{1C} \\
\vdots \\
f_{k(m-2)} \\
f_{k(m-1)} \\
f_{km}
\end{bmatrix}, \quad \quad \quad
\mathbf{b} = \begin{bmatrix}
1 \\
\vdots \\
1 \\
\arrayrulecolor{blue}\hline
0 \\
\vdots \\
0
\end{bmatrix}, \quad \quad \quad
\mathbf{A} \mathbf{x} = \mathbf{b}
\]
}

Vector $\mathbf{x}$ represents each fraction $f_{ij}$ distributed across computer $i$ and map $j$. We now explain the construction of $\mathbf{A}$ and $\mathbf{b}$. The coefficient matrix $\mathbf{A}$ encompasses the two mentioned restrictions, $\mathbf{A}_{eq}$ and $\mathbf{A}_{ub}$, divided by a blue dotted line for clear visualization. Vector $\mathbf{b}$ denotes the bounds for the product of matrix $\mathbf{A}$ and vector $\mathbf{x}$. The top portion of the matrix $\mathbf{A}$ ensures that the total fraction for each computer sums to one by filtering the fractions for each computer using 1 and zeroing out the rest. For example, for the first computer $Comp_{1}$, the first line in matrix $\mathbf{A}$ should consist of $k$ ones, and the rest are zeros. This gives the linear programmer the equation $f_{1A} + f_{1B} + \cdots + f_{1m} = 1$. The top portion of $\mathbf{b}$ consists of ones due to the fraction sum restriction.

The second restriction about approximate TVD equality can also be specified as an equality constraint. However, to provide more flexibility during the linear programming optimization to enable some discrepancies between the overall TVDs if it helps minimize the TVDs further, we design it as an inequality constraint. To manage the flexibility of this inequality constraint, \sol{} includes a tunable parameter, a tolerance rate $\epsilon$. This parameter ensures that TVD values are approximately balanced across different computers within a user-defined bound of $\epsilon$. Suppose $TVD_{i}$ and $TVD_{j}$ represent the overall TVDs of any two random computers $Comp_i$ and $Comp_j$: $TVD_i = f_{iA} \cdot \mathrm{TVD}_{iA} + \cdots + f_{im} \cdot \mathrm{TVD}_{im}$ and $TVD_j = f_{jA} \cdot \mathrm{TVD}_{jA} + \cdots + f_{jm} \cdot \mathrm{TVD}_{jm}$. Any two combinations of computers should meet the following restrictions according to \sol{}'s design:
\begin{equation}
(1 - \epsilon)TVD_i \leq TVD_j \leq (1 + \epsilon)TVD_i \ \ \ \text{and} \ \ \ (1 - \epsilon)TVD_j \leq TVD_i \leq (1 + \epsilon)TVD_j
\end{equation}
\begin{equation}
\text{Combining the two inequalities, we get} \ \ \frac{1}{1 + \epsilon}TVD_i \leq TVD_j \leq (1 + \epsilon)TVD_i
\end{equation}

\begin{equation}
\text{Which can be separated into} \ -TVD_j + \frac{1}{1 + \epsilon}TVD_i \leq 0 \ \ \ \text{and} \ \ \ TVD_j - (1 + \epsilon)TVD_i \leq 0
\label{eq:epsilon}
\end{equation}

This separation ensures that the right side is all zeros, which is reflected in the lower half of $\mathbf{b}$. This inequality constraint is represented by the bottom portion of the matrix $\mathbf{A}$. For any combination of two computers, the sum of $\mathrm{TVD}_{ij}$ multiplied by the corresponding fraction should satisfy Eq.~\ref{eq:epsilon}.

This concludes our description of the setup of \sol{}'s linear programmer. Once this setup is fed into a linear programming engine, it will generate how the shots should be distributed on $Comp_1$ to run the user's circuit. Note that the linear programmer does not need to include all of the computers in the quantum cloud. Any $k$ computers on the cloud can be selected at random. The $k$ computers also do not have to be selected among the currently available computers at the time of circuit submission; only $Comp_1$, which the circuit will actually be run on, has to be available. Other $k-1$ computers can be selected randomly. \textit{This provides stochastic randomization that ensures that the circuit would approximately achieve the same TVD regardless of which computer it is run on, including any computer it may have been run on in the past or any computer that it will be run on in the future.} In our evaluation section (Sec.~\ref{sec:evaluation}), we show that \sol{}'s performance is not highly sensitive to the number of computers chosen beyond $k=2$. \rev{See Appendix~\ref{sec:example} for a worked example.}

\textit{\textbf{However, this raises two questions: (1) how are the circuit maps that are used for the linear programmer selected? And (2) how is the TVD that is fed into the linear programmer for each circuit map estimated? We answer these questions next.}}

\vspace{2mm}

\noindent\textbf{Circuit Maps Generator:} \sol{} generates the maps in a lightweight yet effective manner. Circuit maps can typically be generated in two ways: using the state-of-the-art optimal mapping algorithms available in the Qiskit package~\cite{aleksandrowicz2019qiskit}, which we refer to as OptiMap, or by randomly selecting maps, which we refer to as RandMap. OptiMap includes noise-adaptive algorithms that select qubits with the lowest noise footprint based on the latest calibration data. If we only use maps from OptiMap, they are likely not to be effective. This is because the maps generated by OptiMap are likely to have similar noise properties (similar qubits and similar routing) and thus have similar TVDs, which, although would be estimated to be optimal, do not provide enough diversity for the linear programmer to choose different maps and ensure approximately equal TVDs across different computers. For instance, if all the maps on one computer have a TVD of around $0.3$ and all the maps on another computer have a TVD of around $0.15$, it would be impossible for the linear programmer to select maps that can satisfy a low tolerance (e.g., $\epsilon = 0.1$), as all the maps would be $\approx0.15$ apart. Thus, \sol{} also needs maps in addition to the ones generated by OptiMap.

\sol{}'s RandMap strategy is to employ a ``random walk'' method, which starts with a randomly chosen qubit on the quantum computer. Based on the requirement of the circuit of $q$ qubits, it then connects to $q-1$ adjacent qubits randomly, one after another. \sol{} ensures that all qubits form a connected graph. This way, \sol{} obtains a circuit map of $q$ qubits. \rev{We also generate maps using Qiskit's tools~\cite{aleksandrowicz2019qiskit} transpiler function, which we refer to as OptiMap-based maps since they follow the same approach as the OptiMap baseline. While the Optimap baseline uses only optimization level 3 with no random seed, we configure the transpilation process with randomized seeds and varying optimization levels (0-3) to generate diverse maps.} \textit{In our design, half of the $m$ maps are from $OptiMap$ and the other half are generated using RandMap to provide sufficient diversity in maps to optimize over.} \rev{We adopt this 50/50 split because OptiMap typically yields lower-TVD maps but can be unstable across calibration cycles, whereas RandMap provides diverse alternatives that sometimes play a crucial role in giving the linear programmer options to equalize TVD across computers and days. 50/50 consistently yields balanced performance and stability.} \sol{}'s linear programmer would distribute shots across these $m$ possible maps. In our evaluation section (Sec.~\ref{sec:evaluation}), we demonstrate the impact of the number of maps chosen, $m$, on the performance of \sol{}.


Next, we discuss how \sol{} estimates the TVD of each of these maps.

\vspace{2mm}

\noindent\textbf{The \sol{} TVD Predictor:} In an ideal yet impractical scenario, one could run each map on every computer to obtain precise TVD measurements and use this information to distribute the shots. However, this approach is not feasible due to the significant time and resource constraints involved. In such scenarios, the estimation of the output error of a quantum circuit is widely done using a metric called Estimated Success Probability (ESP)~\cite{patel2023graphine,li2022optimal,brandhofer2023optimal,tannu2019ensemble,xie2021mitigating,patel2022geyser,ludmir2024pachinqo,ludmir2024parallax}. ESP is calculated by multiplying the success probabilities of each gate, which is based on the gate error rates and the number of gates. Thus, it can be obtained simply from the calibration data without running the circuit. However, this metric has two shortcomings for the context of \sol{}: (1) ESP only incorporates data from a single calibration cycle, failing to account for noise variance over time. As a result, using it will not enable \sol{} to tackle temporal variations (even though the linear programmer tackles spatial variations using multiple computers). (2) Our empirical results (illustrated in Sec.~\ref{sec:evaluation}) found this metric to be a poor estimator of the TVD for superconducting systems as it does not take into account complex noise interactions. Thus, a critical challenge for \sol{} is to be able to estimate the TVD of a circuit on a given map without running it first.

To address this, \sol{} leverages a machine-learning TVD predictor built using Random Forest~\cite{breiman2001random} to estimate the TVD for each circuit transpiled to a circuit map. This approach allows us to predict TVD without the need for extensive real-world measurements more accurately. \textit{Further, \sol{}'s TVD predictor is trained on data from numerous calibration cycles, enabling it to account for noise variance and thereby provide more stable and accurate predictions over time.} \sol{} introduces an innovative approach to predicting TVD given a circuit transpiled to a map. It employs a two-number representation to prepare a circuit as input for the random forest model. The circuit is decomposed into a sequence of only four types of basis gates: X gates, SX gates, CX gates, and RZ gates. These are the basis gates on IBM hardware~\cite{castelvecchi2017ibm}, but \sol{} can function with any other basis set as well. Since Rz gates are virtual and noise-free, only X, SX, and CX gates are considered in \sol{}'s TVD learning for circuits. Each gate is represented as a pair of two numbers, which we call a two-number representation. For one-qubit gates (X and SX), a gate is represented as $(q_i, 0)$ where the $q_i$ indicates the gate operates on the $q_i^{th}$ qubit. For two-qubit CX gates, the representation is $(q_c, q_t)$ where $q_c$ is the control qubit index and $q_t$ is the target qubit index. 

\begin{figure}[t]
    \centering
    \includegraphics[scale=0.3105]{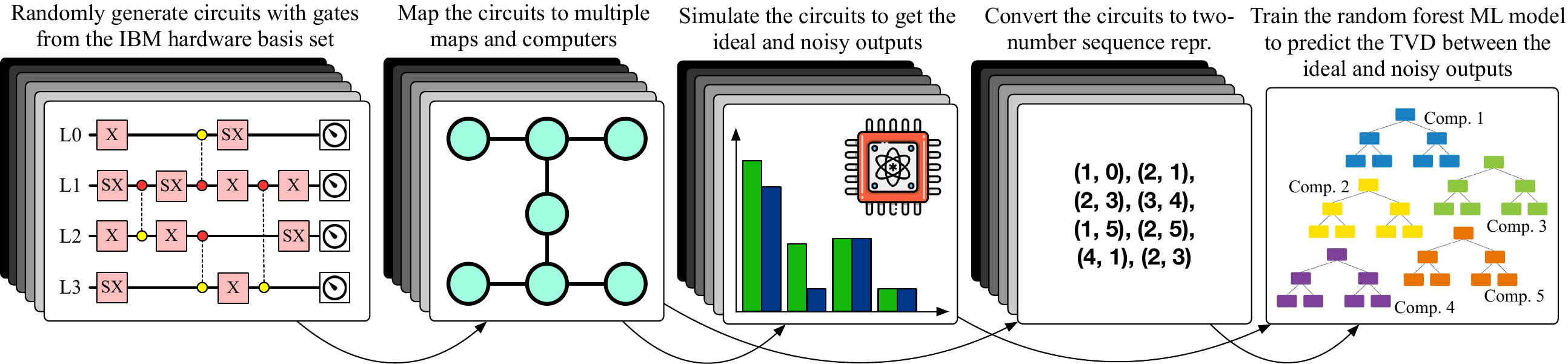}
    \vspace{1mm}
    \hrule
    \vspace{-3.5mm}
    \caption{\sol{}'s TVD Predictor trains random forest models using self-generated random circuits composed of X, SX, and CX basis gates, represented as two-number sequences, to learn TVD.}
    \label{fig:training}
    \vspace{-4mm}
\end{figure}

As shown in Fig~\ref{fig:training}, \sol{} generates the training dataset for the random forest model from randomly generated circuits consisting of combinations of X, SX, and CX gates. These circuits are first transpiled to random maps and then converted into the two-number representation sequence, which is subsequently flattened. \sol{} uses a Random Forest Regressor~\cite{breiman2001random}, constructing multiple decision trees during the training phase and outputting the mean prediction of individual trees. Given the variability in sequence lengths for circuits, each sequence is padded with pairs of zeros to ensure uniform length across the dataset. To enhance model performance and ensure comparable feature scales, the data is normalized with a mean of zero and a standard deviation of one. After preprocessing, the normalized, padded, and flattened sequences are used as input features for training the RandomForest Regressor. 

Due to the significant differences in TVD introduced by different noise models, each computer is provided with its own separate model. This enables \sol{} to train for the noise behavior of each computer, taking into account noise variations over multiple days as the training data is collected over multiple days. However, \sol{} does not use this timing information as an input feature to ensure that the model learns generalized noise behavior and predicts TVDs for maps based on their generalized performance across different times and not for a particular time. \textit{This key decision enables \sol{} to reduce temporal variability in multiple runs of the same circuit.} \rev{The use of the random forest model also enables \sol{} to make high-confidence predictions: the list of predicted TVDs generated by each tree is processed through a random forest confidence filter. This filter selects the top $\frac{2}{3}$ predicted TVD values with the lowest variance among the decision trees within the random forest ensemble. Retaining all tree predictions would include unstable outliers from high-variance trees, while keeping too few predictions reduces ensemble diversity and introduces bias by over-representing a narrow subset of the forest. Among several evaluated retention thresholds, $\frac{2}{3}$ achieved the lowest mean squared error in TVD prediction.} By doing so, we ensure that only the most reliable predictions are used in order to get optimal results from the linear programmer. 

\begin{figure}[t]
    \centering
    \includegraphics[scale=0.4]{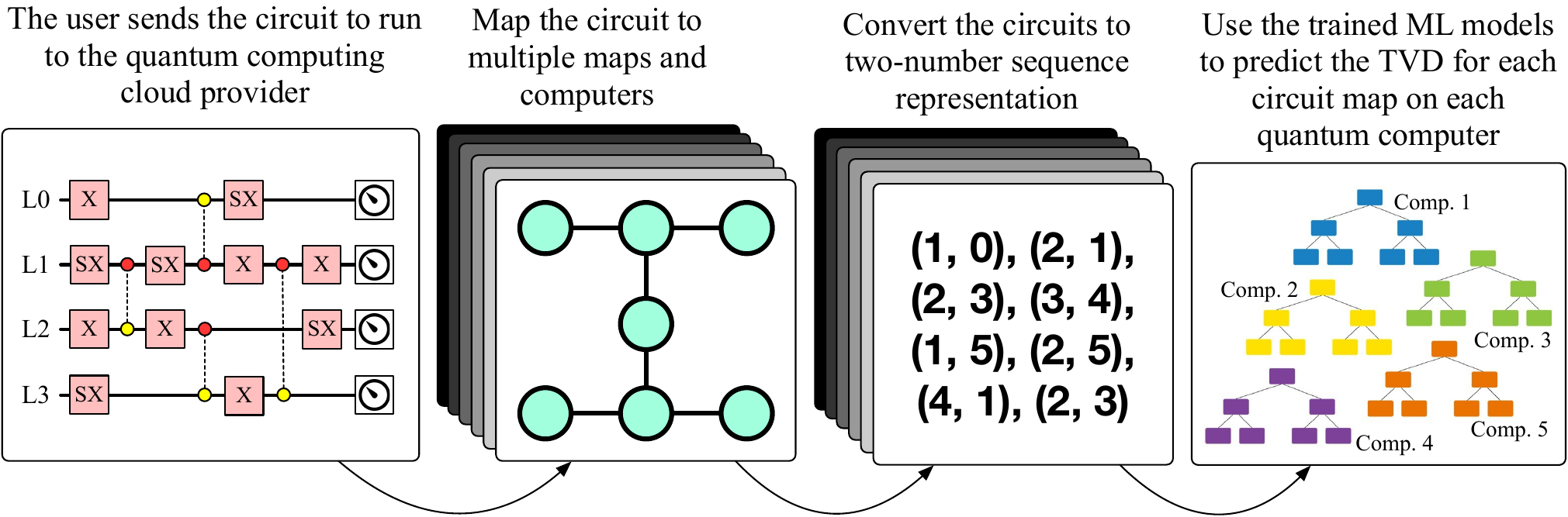}
    \vspace{1mm}
    \hrule
    \vspace{-3.5mm}
    \caption{\sol{} transforms the target circuit into a two-number representation to predict its TVD.}
    \label{fig:inference}
    \vspace{-4mm}
\end{figure} 

The prediction or inference process follows the approach depicted in Fig~\ref{fig:inference}. It involves the same two-number representation for the three basis gates (X, SX, and CX). For each computer, the corresponding model is used to infer the TVD upon the circuit. \textit{It is important to note that only the quantum circuit is used as an input to the model; no other feature that relies on the running of the circuits is used.} Note also that simply predicting the TVD does not eliminate the need to run the circuit eventually after running \sol{} linear programmer, as the circuit's output distribution is still completely unknown. Only the potential overall error in the output is predicted.

This concludes our discussion of \sol{}'s design. Next, we present the evaluation methodology, followed by our experimental results.

%% file: sections/methodology.tex
\section{Methodology}
\label{sec:methodology}

\noindent\textbf{Experimental Setup:} We utilized IBM's Qiskit software development kit \cite{aleksandrowiczqiskit} for all experiments. Simulations were performed using the $AerSimulator$ from Qiskit's Aer module on five IBM quantum computers: $ibm\_cairo$, $ibm\_hanoi$, $ibm\_algiers$, $ibmq\_mumbai$, and $ibmq\_kolkata$, each equipped with 27 physical qubits. Noise models were derived from these quantum computers over multiple days. Conducting experiments on real quantum hardware for all runs was impractical due to the extensive number of runs required. Therefore, in keeping with the prior work~\cite{lao2021designing,liu2020reliability,maurya2023scaling,murali2020software,jin2024tetris}, we largely relied on noisy simulations using the daily-updated noise models of the real IBM quantum computers. \rev{Nontheless, we did perform validation experiments on real computers as well: \textit{ibm\_sherbrooke}, \textit{ibm\_kyiv}, \textit{ibm\_brisbane}, \textit{ibm\_brussels}, and \textit{ibm\_strasbourg}.}

\rev{We note that the noise models used in our study are generated and provided directly by IBM Quantum, and they faithfully emulate the noise characteristics of the real hardware. Each model is derived from daily calibration runs that benchmark qubit properties, ensuring that the model reflects temporally varying device conditions~\cite{huo2026nest,huo2025revisiting,han2025enqode}. Specifically, the noise models capture properties for each qubit, including average $\overline{T_1}$ and $\overline{T_2}$ coherence times, single-qubit and two-qubit gate durations, single-qubit error rates, two-qubit error rates, and the qubit connectivity map. These error rates fall in the realistic NISQ regime of $10^{-1}$–$10^{-3}$ that current superconducting devices operate under. Importantly, our real-hardware experiments are run on a different set of IBM quantum computers for which we collected these daily noise models, ensuring results across between simulated noise characteristics and actual device behavior.}

For our study, we evaluated 20 different algorithms, each requiring evaluation of 4 distinct techniques with 100 executions per technique, resulting in a total of 8,000 simulation runs. The hardware configuration included an NVIDIA GeForce RTX 3060 Ti GPU for machine learning task training and a 12th Gen Intel(R) Core(TM) i5-12600KF 3.70 GHz CPU with 32.0 GB RAM, running on Windows 11. The software environment comprised Python 3.11.7 and Qiskit 0.46.0.

\vspace{2mm}

\noindent\textbf{TVD Predictor Implementation:} The training dataset comprised exclusively randomly generated circuits, omitting the algorithms used for evaluation to ensure independent performance testing for those algorithms. Each experiment involved 100 runs per algorithm. A total of 289,107 random circuits were generated to train the Random Forest predictor. The number of gates in the training data ranged from 2 to 655 after transpilation. Each circuit, both for optimal maps and random maps, was routed according to its specific noise model using an optimization level of three. Random Forest implementation was done using $RandomForestRegressor$ from $sklearn.ensemble$. \sol{} used 100 trees with a maximum depth of 10. The minimum number of samples required to split an internal node was set to 2, and at least 1 sample was required at a leaf node. The number of features considered at each split was set to the square root of the total features. Bootstrap samples were employed, and the random state was set to 42 for reproducibility. Using this configuration, we achieved a low mean squared error (MSE) of 0.04 between the actual TVD and the predicted TVD across all 20 algorithms, and each was tested 100 times. We also experimented with using Long Short-Term Memory (LSTM) networks~\cite{hochreiter1997long} and attention-based transformer networks~\cite{vaswani2017attention} for prediction, as they are designed to learn from sequential data, and the gate sequences in circuits might exhibit similar attributes. However, the results were not optimal, possibly due to the complexity of these models not being well-suited to our data set, leading to overfitting.

\vspace{2mm}

\noindent \rev{\textbf{TVD Predictor's Robustness to Hardware Drift.} Our TVD predictors are trained using large, randomly structured quantum gate sequences with different noise characterizations across multiple computers over days. This approach enables the TVD predictor to capture natural drift patterns rather than instantaneous calibration snapshots. The temporal diversity in this training methodology enhances the predictor's ability to generalize beyond the specific calibration cycles observed during training. When hardware drift causes the predicted TVD for any individual map to become inaccurate, this distributed approach limits the overall adverse performance impact.} 

\vspace{2mm}

\noindent\textbf{Linear Programmer Implementation:} We used SciPy 1.12.0 for linear programming tasks. Equality constraints were implemented using SciPy's $linprog$ function. Inequality constraints of the form $\mathbf{A}_{ub} \mathbf{x} \leq \mathbf{b}_{ub}$ were transformed into $\mathbf{A}_{ub} \mathbf{x} - \mathbf{b}_{ub} \leq 0$ to align with the interface requirements of $linprog$. Python arrays were used for matrix representation. For uncertainty filtering, we identified and removed the lowest $\frac{1}{3}$ of the variance predictions among the decision trees within the random forest ensemble. This involved checking the predicted TVDs for the target computer, filtering the results, and removing other TVDs across all computers according to the filtered index before proceeding to the linear programming stage.

\vspace{2mm}

\begin{table}[t]
    \centering
    \caption{List of algorithms and benchmarks used to evaluate \sol{}. Their primary properties are also listed. ``\#1Q'' refers to the numbers of one-qubit gates and ``\#2/3+Q'' refers to the number of multi-qubit gates.}
    \vspace{-3mm}
    \scalebox{0.83}{
    \begin{tabular}{>{\columncolor{cyan!10}}c>{\columncolor{pink!10}}lcccc}
    \hline
    \textbf{Algo.} & \textbf{Description} & \textbf{\#Qubits} & \textbf{\#1Q Gates} & \textbf{\#2/3+Q Gates} & \textbf{Depth} \\
    \hline
    BC & Basis Change Circuit & 3 & 23 & 10 & 22 \\
    FRD & Controlled-Swap Gate Circuit & 3 & 11 & 8 & 12 \\
    LNS & Linear Solver Circuit of One Qubit & 3 & 9 & 4 & 10 \\
    QAOA & Quantum Approximate Optimization Algo. & 3 & 9 & 6 & 12 \\
    TEL & Quantum Teleportation Circuit & 3 & 6 & 2 & 7 \\
    TOF & Toffoli Gate Circuit & 3 & 12 & 6 & 13 \\
    ADD & Quantum Ripple-Carry Adder Circuit & 4 & 13 & 10 & 12 \\
    BELL & Bell State Circuit & 4 & 12 & 7 & 10 \\
    CAT & Cat State Circuit & 4 & 1 & 3 & 5 \\
    HS4 & Hidden Subgroup Problem & 4 & 24 & 4 & 10 \\
    IQFT & Inverse Quantum Fourier Transform & 4 & 14 & 0 & 12 \\
    QFT & Quantum Fourier Transform & 4 & 6 & 6 & 9 \\
    VAR & Variational Quantum Circuit & 4 & 38 & 16 & 34 \\
    VQE & Variational Quantum Eigensolver & 4 & 64 & 9 & 24 \\
    LNP & Learning Parity with Noise & 5 & 9 & 2 & 5 \\
    QEC & Quantum Error Correction Encoding & 5 & 15 & 10 & 18 \\
    SIM & Simon's Algorithm & 6 & 12 & 2/2 & 9 \\
    QAOA6 & Quantum Approximate Optimization Algo. & 6 & 92 & 54 & 66 \\
    QPE & Quantum Phase Estimation & 9 & 15 & 16/2 & 21 \\
    ISNG & Ising Model Simulation & 10 & 390 & 90 & 71 \\
    \hline
    \end{tabular}}
    \vspace{-3mm}
    \label{tab:algos}
\end{table}

\noindent\textbf{Quantum Algorithms and Benchmarks:} Table~\ref{tab:algos} list the algorithms used for \sol{}'s evaluation and their properties. The benchmark that we use is QASMBench~\cite{li2023qasmbench}. We use algorithms that cover different characteristics to ensure they have a variety of output behaviors and are differently impacted by hardware noise. We made sure that the algorithms are as large and deep as can be given two requirements: (1) they have to be classically simulatable so that we can get their ideal output to calculate the TVD against to study the output fidelity on noisy computers, and (2) they cannot be too deep; otherwise, they would get considerably impacted by noise such that their output TVD would reach beyond 50\%, resulting in uninterpretable output.

\vspace{2mm}

\noindent\textbf{Competitive Techniques:} We evaluate \sol{} against three other techniques: \textbf{(1) OptiMap:} This technique selects the optimal circuit map for an algorithm using all noise-adaptive optimizations to a circuit on a given quantum computer based on the noise information available from the current calibration cycle~\cite{aleksandrowiczqiskit}. All shots are run on the optimal circuit map. \textbf{(2) RandMap:} The technique selects any random map on a given quantum computer to run the circuit. All shots are run on this random map. OptiMap and RandMap are routed by using Qiskit optimization level 3. \textbf{(3) EqualDist:} The technique selects the same maps as \sol{} but divides the shots equally among all the maps instead of optimizing the number of shots using the linear programmer. This will help assess the efficacy of \sol{}'s linear programmer. All techniques run the same number of total shots for a given circuit: 32,000 (same as the total number of shots used for \sol{}).

\vspace{2mm}

\noindent\textbf{Evaluation Metrics:} \textbf{(1) Mean TVD \textit{(lower is better)}:} This is the average TVD when the same quantum circuit is submitted for running multiple times (including across different computers and over multiple days). This metric helps examine the average output quality. \textbf{(2) TVD Std. Dev. \textit{(lower is better)}:} This is the coefficient of variation (standard deviation divided by the mean) of TVD of the same quantum circuit over multiple runs. This metric helps examine the variation in the output quality. \textbf{(3) Mapping Time \textit{(lower is better)}:} This metric measures the overall online time required to map a given quantum circuit to the quantum hardware for different techniques. Note: \sol{} is the only technique with a one-time offline time overhead of training the TVD predictor. The actual training took only $\approx$10 minutes in total for all five computers. The training data collection took around a week to ensure coverage of temporally varying noise characteristics. All competitive techniques (including \sol{}) have online time overheads.

%% file: sections/evaluation.tex
\section{Evaluation and Analysis}
\label{sec:evaluation}

In this section, we discuss the evaluation results of \sol{}.

\vspace{2mm}

\begin{figure}
    \centering
    \includegraphics[scale=0.765]{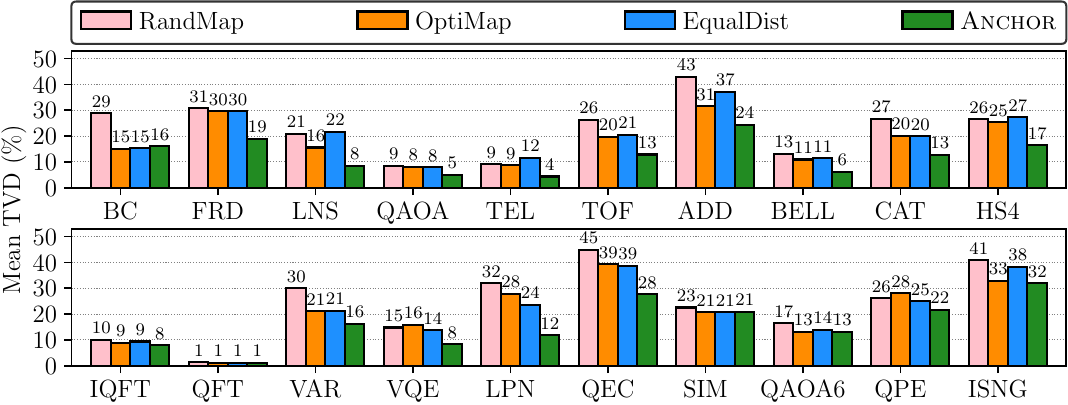}
    \vspace{1mm}
    \hrule
    \vspace{-3.5mm}
    \caption{\sol{} achieves a lower mean TVD on average as compared to other competitive techniques.}
    \label{fig:main_mean_tvd}
    \vspace{-3mm}
\end{figure}

\begin{figure}
    \centering
    \includegraphics[scale=0.765]{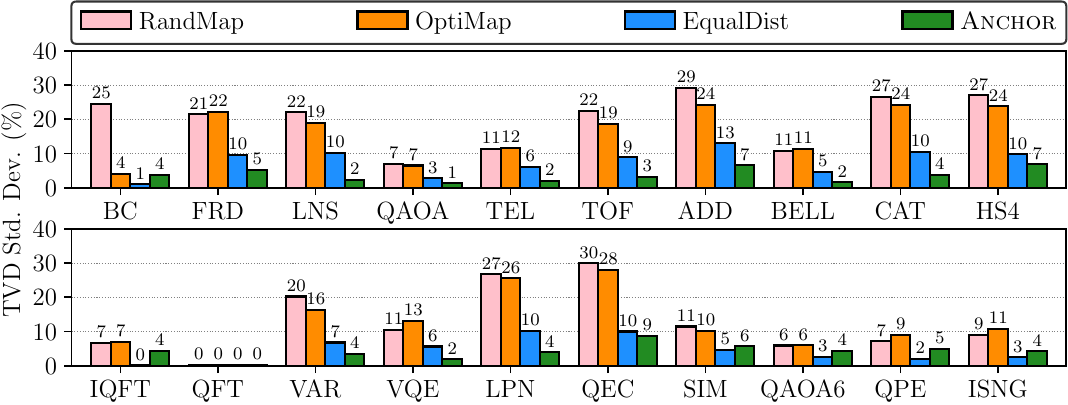}
    \vspace{1mm}
    \hrule
    \vspace{-3.5mm}
    \caption{\sol{} achieves a lower TVD standard deviation on average as compared to other techniques.}
    \label{fig:main_tvd_std_dev}
    \vspace{-5mm}
\end{figure}

\noindent\textbf{Reduction in the Overall TVD Variability while Maintaining a Low Mean TVD.} \textit{The analysis of the mean TVD and TVD standard deviation (Fig.~\ref{fig:main_mean_tvd} and Fig.~\ref{fig:main_tvd_std_dev}) shows that \sol{} outperforms the other mapping techniques across various quantum algorithms and benchmarks.} In both metrics, lower values indicate better performance, as they reflect a smaller deviation from the ideal output of an algorithm when executed on noisy quantum computers.

Regarding the mean TVD (Fig~\ref{fig:main_mean_tvd}), \sol{} consistently achieves lower TVD values across the majority of algorithms. For instance, in the TOF algorithm, \sol{} achieves a Mean TVD of 13\%, significantly lower than RandMap (26\%), OptiMap (20\%), and EqualDist (21\%). This trend is observed across other algorithms such as QAOA, TEL, and QEC, where \sol{} maintains the lowest mean TVD values, illustrating its effectiveness in reducing errors. On average, \sol{} shows a reduction in mean TVD of 39\% over RandMap, 27\% over OptiMap, and 30\% over EqualDist.

As expected, RandMap has the highest TVD in general. A surprising result here is in the case of OptiMap, which uses all of Qiskit's noise-adaptive optimizations and, thus, is expected to have one of the lowest mean TVDs. While it often performs better than RandMap, it still falls short compared to \sol{}. This indicates that while noise-adaptive optimizations can improve performance, they are not as effective as \sol{}'s linear programming approach that considers behaviors across multiple computers and time periods, which optimizes shot distribution based on noise variability. The reason for this is the noise models that these noise-adaptive optimizations rely on often are not reliable even if they are generated during the same calibration cycle as the algorithm runs. This is because qubits' noise properties change frequently and stochastically, which greatly reduces the effectiveness of the optimizations. EqualDist, which uses the same maps as \sol{} but without optimized shot distribution, generally performs better than RandMap but worse than \sol{}, underscoring the importance of \sol{}'s shot optimization. The performance of EqualDist demonstrates that the selection of maps alone is insufficient; the distribution of shots among these maps plays a critical role in minimizing both the mean TVD and its variability.

\begin{figure}
    \centering
    \subfloat[Temporal Variability]{\includegraphics[scale=0.47]{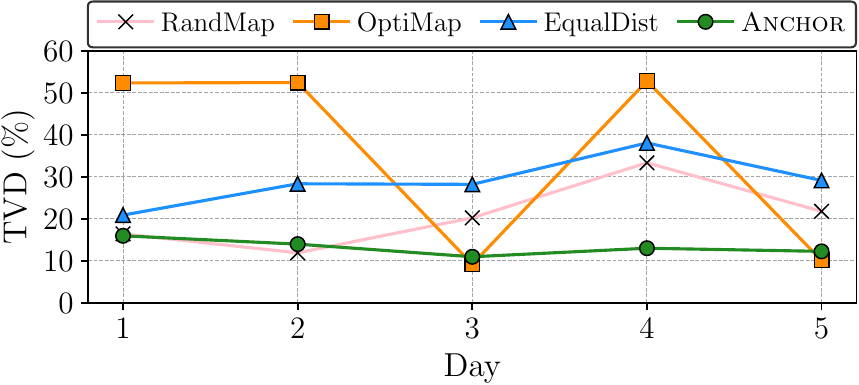}}
    \hfill
    \subfloat[Spatial Variability]{\includegraphics[scale=0.47]{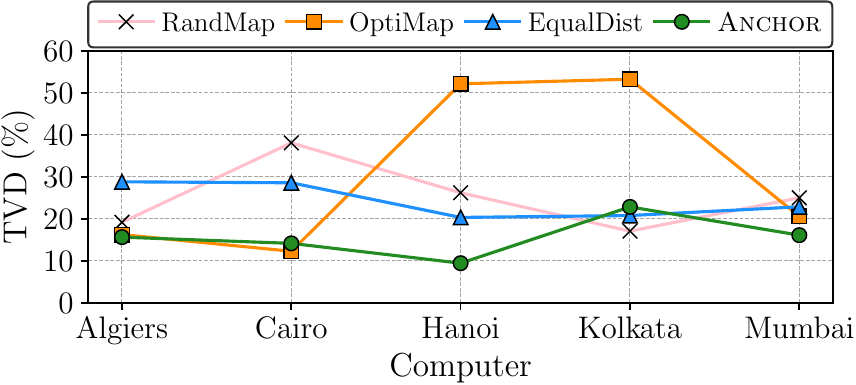}}
    \vspace{1mm}
    \hrule
    \vspace{-3.5mm}
    \caption{(a) Temporal and (b) spatial variability in the TVD of a quantum circuit when the 4-qubit VAR algorithm is run on the IBM quantum computers shows the degree by which \sol{} stabilizes the results.}
    \label{fig:temp_spa}
    \vspace{-4mm}
\end{figure}

In the case of TVD standard deviation, Fig.\ref{fig:main_tvd_std_dev} highlights the robustness of \sol{} in minimizing performance variability. For example, for the TOF algorithm, \sol{} achieves a TVD standard deviation of approximately 3\%, compared to 22\% for RandMap, 19\% for OptiMap, and 9\% for EqualDist. For some shallow algorithms with low TVD, EqualDist is able to achieve similar or slightly lower TVD than \sol{} (e.g., QAOA6, ISNG).  
This is because EqualDist’s uniform shot distribution might sometimes incidentally align better with the noise characteristics of some algorithms, resulting in slight improvement over \sol{}. Nonetheless, \sol{} achieves a lower mean TVD for these algorithms. On average, \sol{} shows a reduction in TVD standard deviation of 76\% over RandMap, 73\% over OptiMap, and 45\% over EqualDist.

This significant reduction in variability demonstrates that \sol{} reduces the mean error and ensures more consistent performance across different runs. This consistency is crucial for practical quantum computing applications, where reliability is a key concern. The lower performance variability helps ensure that the output fidelity of quantum circuits is more stable, regardless of temporal fluctuations or spatial differences across quantum computers.

\vspace{2mm}

\noindent\textbf{Reduction in Temporal and Spatial Variability.} We now explicitly evaluate how TVD varies across different days and computers (similar to the analysis in Fig.~\ref{fig:motiv} in Sec.~\ref{sec:motivation}, but with additional data of EqualDist and \sol{}). In Fig.~\ref{fig:temp_spa}(a), which depicts the TVD of the VAR algorithm over five days on the IBM Hanoi quantum computer, \sol{} consistently achieves the lowest TVD values, maintaining stability around 12-15\%, compared to RandMap and OptiMap, which exhibit higher variability, often exceeding 20\%. EqualDist shows moderate variability but is still higher than \sol{}. This indicates that \sol{}'s learning-based TVD predictor effectively mitigates temporal noise fluctuations, ensuring more reliable and consistent results over time.

In Figure Fig.~\ref{fig:temp_spa}(b), which shows the TVD of VAR across different IBM quantum computers (Algiers, Cairo, Hanoi, Kolkata, and Mumbai) on the same day, \sol{} again outperforms other techniques, maintaining TVD values around 10-20\%, while RandMap and OptiMap frequently reach 30-40\%. EqualDist performs better than RandMap and OptiMap but still falls short of \sol{}. This consistent performance highlights \sol{}'s ability to account for spatial noise variability, delivering more stable and predictable results regardless of the quantum computer used. Overall, these figures validate \sol{}'s effectiveness in stabilizing quantum circuit performance by addressing both temporal and spatial variability, significantly outperforming existing methods.

\vspace{2mm}

\begin{table}[t]
    \centering
    \caption{\sol{} can have a higher but manageable online mapping time than other comparative techniques -- all times are listed in seconds. This is mainly due to the linear programming component of \sol{}.}
    \vspace{-3mm}
    \scalebox{0.83}{
    \begin{tabular}{>{\columncolor{cyan!10}}ccccccccccc}
    \hline
    \textbf{Technique} & BC & FRD & LNS & QAOA & TEL & TOF & ADD & BELL & CAT & HS4 \\
    \hline
    RandMap & 2.2 & 2.01 & 1.89 & 2.13 & 1.81 & 1.88 & 1.95 & 1.83 & 1.71 & 1.84 \\
    OptiMap & 0.35 & 0.38 & 0.33 & 0.35 & 0.47 & 0.53 & 0.61 & 0.53 & 0.59 & 0.52 \\
    EqualDist & 0.34 & 0.32 & 0.34 & 0.48 & 0.48 & 0.49 & 0.51 & 0.50 & 0.55 & 0.46 \\
    \sol{} & 7.45 & 6.65 & 3.83 & 4.54 & 7.82 & 11.97 & 17.48 & 14.41 & 11.77 & 11.90 \\
    \hline
    \hline
    \textbf{Technique} & IQFT & QFT & VAR & VQE & LPN & QEC & SIM & QAOA6 & QPE & ISNG \\
    \hline
    RandMap & 1.85 & 1.95 & 1.83 & 1.85 & 1.74 & 1.99 & 2.21 & 3.08 & 3.23 & 3.52 \\
    OptiMap & 0.51 & 0.62 & 0.52 & 0.54 & 0.49 & 0.58 & 0.64 & 1.08 & 0.64 & 0.55 \\
    EqualDist & 0.50 & 0.51 & 0.48 & 0.50 & 0.47 & 0.52 & 0.52 & 0.76 & 0.79 & 5.74 \\
    \sol{} & 9.41 & 17.91 & 15.85 & 20.85 & 8.82 & 18.01 & 64.70 & 69.20 & 126.52 & 29.70 \\
    \hline
    \end{tabular}}
    \vspace{-3mm}
    \label{tab:map_times}
\end{table}

\noindent\textbf{Mapping Times of Different Techniques.} While \sol{} demonstrates significant improvements in reducing both temporal and spatial variability in the TVD of quantum circuits, it is important to acknowledge that the mapping time for \sol{} is higher compared to other techniques (shown in Table~\ref{tab:map_times}). The average mapping time of RandMap is 2.13 seconds, OptiMap is 0.54 seconds, EqualDist is 0.76 seconds, and \sol{} is 23.94 seconds. OptiMap has the lowest overhead, as it requires only one transpilation. EqualDist takes slightly more time because it performs transpilation for multiple maps. RandMap has a higher overhead since it involves traversing the qubit connectivity graph to create a random map. \rev{\sol{}'s mapping time overhead is entirely classical. It stems from transpiling maps with various optimization levels and generating RandMap, followed by applying linear programming with a learning-based TVD predictor to optimize shot distribution across these maps while accounting for complex noise interactions. The linear programming component is computationally efficient, typically completing within 1–2 seconds, while the majority of the overhead arises from on-demand map generation and transpilation.}

\rev{Nonetheless, this overhead remains tractable and can be further reduced through lightweight optimizations. For larger circuits, pre-generating and caching a library of random maps eliminates repeated on-the-fly generation. Additionally, generating OptiMap-based maps from multiple transpilation levels is inherently parallelizable across classical cores. Given that \sol{} achieves measurable improvements in circuit fidelity and reduced performance variability, the modest classical overhead represents a favorable trade-off for practical quantum applications.}


\vspace{2mm}

\begin{figure}
    \centering
    \subfloat[Mean TVD]{\includegraphics[scale=0.69]{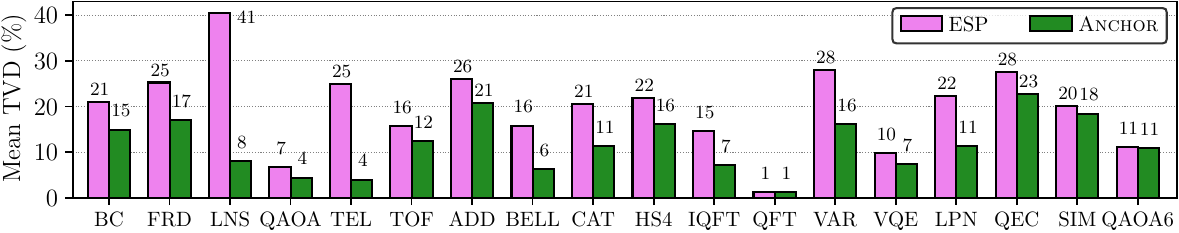}}
    \vspace{-4.5mm}
    \subfloat[TVD Standard Deviation]{\includegraphics[scale=0.69]{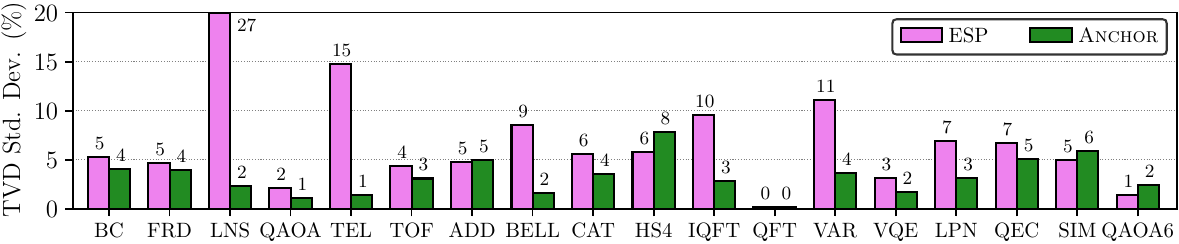}}
    \vspace{1mm}
    \hrule
    \vspace{-3.5mm}
    \caption{When comparing \sol{}'s linear programmer implemented with ESP (labeled ESP) and its linear programmer implemented with its TVD predictor (labeled \sol{}), \sol{}'s TVD predictor-based implementation outperforms ESP-based implementation considerably.}
    \label{fig:esp}
    \vspace{-4mm}
\end{figure}


\noindent\textbf{Using ESP instead of \sol{}'s TVD Predictor:} To evaluate the significance of \sol{}'s TVD predictor, we compare \sol{} against its linear programmer being implemented with the ESP metric. Fig.~\ref{fig:esp}(a) and Fig.~\ref{fig:esp}(b) show the results for mean TVD and TVD standard deviation. Note: due to the inaccuracy of ESP, it estimated 100\% error for some of the circuits, which made the linear programmer inapplicable for them -- therefore, we had to omit those circuits for this analysis.

\sol{} consistently achieves lower mean TVD and TVD standard deviation values, indicating more reliable and stable performance. For instance, in the BELL algorithm, \sol{} achieves a TVD standard deviation of approximately 2\%, significantly lower than the 9\% observed for ESP. Similar trends are observed for algorithms like LNS and VAR, where \sol{} maintains TVD Standard Deviation values around 2\% and 4\%, respectively, compared to ESP's 27\% and 11\%. These results demonstrate that \sol{}'s learning-based TVD predictor effectively accounts for complex noise interactions, providing more accurate and consistent estimations of performance variability.

The discrepancies between ESP and \sol{} are most pronounced in more complex algorithms, underscoring the limitations of ESP in capturing intricate noise dynamics in superconducting systems. While ESP calculates success probabilities based on individual gate success rates and coherence times at a given time, it fails to consider the combined noise interactions, leading to higher variability. In contrast, \sol{}'s TVD predictor, trained to understand these interactions for dynamic noise conditions, significantly reduces variability. Overall, the results validate the efficacy of \sol{}'s approach, justifying the offline overhead of training the TVD predictor and highlighting the inadequacy of ESP as a reliable noise estimator in quantum computing.

Next, we ablate \sol{}'s performance with respect to several design decisions.





\vspace{2mm}

\begin{figure}
    \centering
    \includegraphics[scale=0.765]{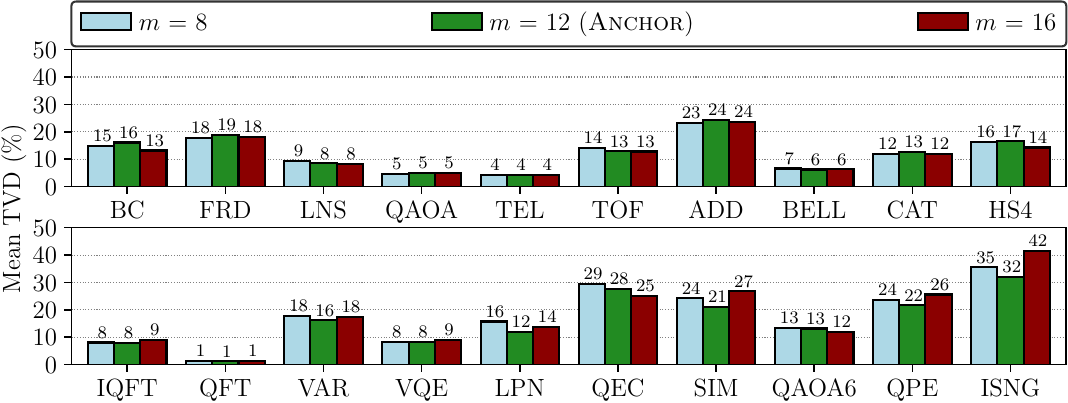}
    \vspace{1mm}
    \hrule
    \vspace{-3.5mm}
    \caption{The mean TVD is largely unaffected by the choice of the number of maps used in the optimization.}
    \label{fig:map_mean}
    \vspace{-4mm}
\end{figure}

\begin{figure}
    \centering
    \includegraphics[scale=0.765]{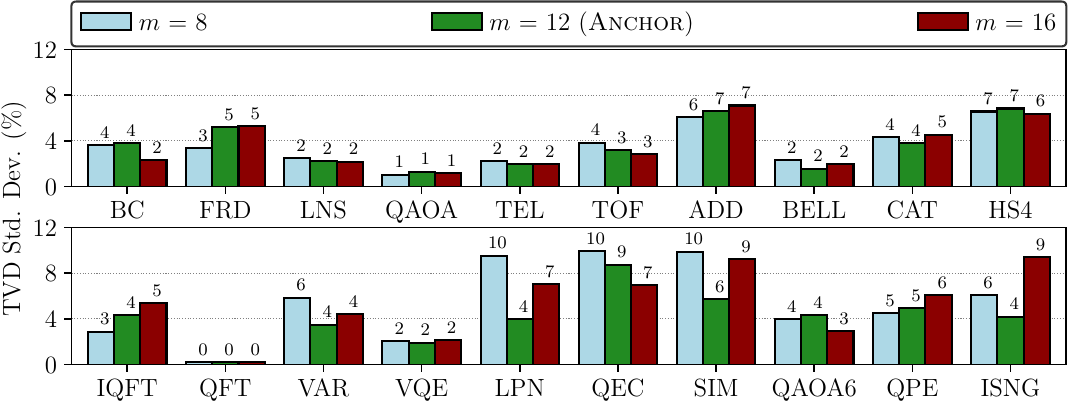}
    \vspace{1mm}
    \hrule
    \vspace{-3.5mm}
    \caption{The TVD standard deviation also remains mostly unchanged, but using 12 maps performs the best for some algorithms compared to using 8 or 16 maps (e.g., VAR, SIM, ISNG).}
    \label{fig:map_std_dev}
    \vspace{-4mm}
\end{figure}

\noindent\textbf{Ablation Study on \sol{}'s Optimization Parameters:} We evaluate the impact of three key parameters in \sol{}'s optimization.  First, the number of maps used in the optimization process ($m$) has a negligible impact on mean TVD but plays a crucial role in controlling performance variability. Fig.~\ref{fig:map_mean} and Fig.~\ref{fig:map_std_dev} analyze the impact of the number of maps used in the optimization of \sol{}'s linear programmer ($m$) on the mean TVD and TVD standard deviation, respectively. Across benchmarks such as ADD, mean TVD remains relatively stable (within 1\%) across values of $m = 8, 12,$ and $16$. However, TVD standard deviation trends indicate that using too few maps (e.g., $m = 8$) limits \sol{}'s ability to distribute computational loads effectively, resulting in variability increases of up to 4\% in some algorithms. Conversely, using too many maps (e.g., $m = 16$) can introduce noise due to optimization complexity, leading to increased variability in benchmarks such as SIM and ISNG. Empirical results show that using $m = 12$ provides an optimal balance, reducing the variability by 3 to 4\% compared to other choices.

Second, varying the number of quantum computers used in \sol{}'s linear programming optimization impacts performance stability. While we do not plot these results for brevity, our analysis is backed by empirical evaluations across all quantum benchmarks listed in Table~\ref{tab:algos}. Across algorithms such as LPN, QAOA, and TEL, mean TVD differences between using 2, 3, or 4 computers remain within 1–2\%. However, TVD standard deviation shows noticeable fluctuations, with larger quantum workloads such as QPE and ISNG exhibiting increased variability when using fewer computers. Using four quantum computers results in more stable performance with an observed variability reduction of up to 3\% in some benchmarks. While adding more computers increases the runtime of the linear programmer due to additional optimization constraints, the overhead remains negligible ($<1$ second), making this trade-off favorable for improved stability.

Last, adjusting the tolerance parameter ($\epsilon$) directly affects the trade-off between minimizing mean TVD and maintaining stability. A lower $\epsilon$ (e.g., 0.01) limits variability but reduces \sol{}'s ability to effectively minimize mean TVD (e.g, algorithms like ISNG exhibit a mean TVD of 39\% at $\epsilon = 0.01$, which drops to 32\% at $\epsilon = 0.1$). Conversely, higher tolerance levels (e.g., $\epsilon = 0.5$) primarily focus on reducing mean TVD but can lead to increased performance variability. \sol{}'s selected level of $\epsilon = 0.1$ balances mean TVD reduction while keeping variability manageable across diverse quantum workloads.

\vspace{2mm}

\begin{figure}[t]
    \centering
    \includegraphics[width=0.99\columnwidth]{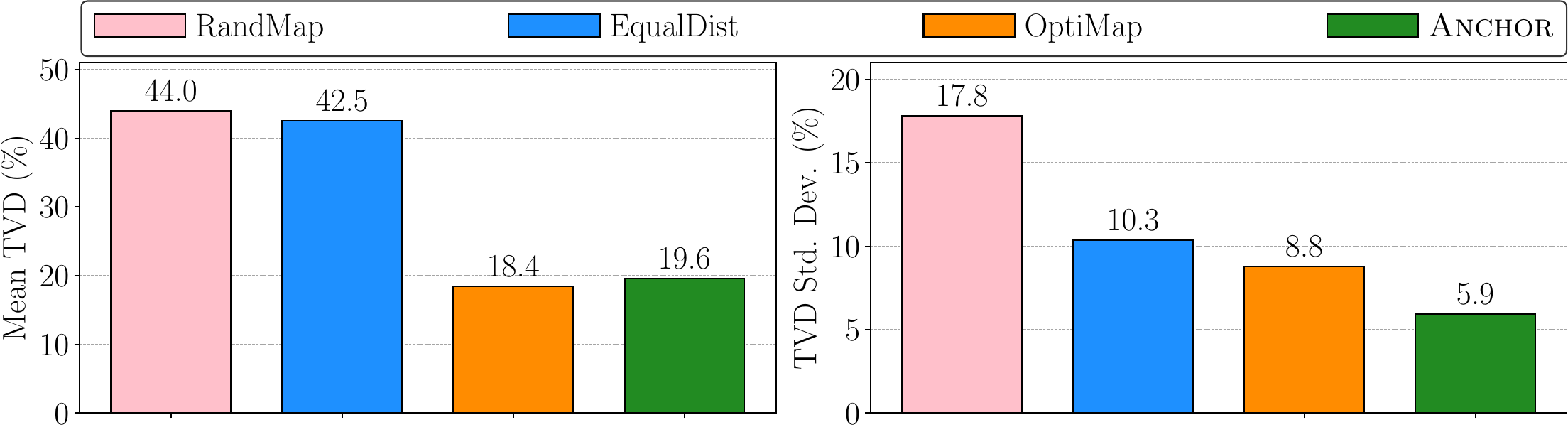}
    \vspace{1mm}
    \hrule
    \vspace{-3.5mm}
    \caption{\sol{} achieves the lowest TVD standard deviation across five IBM quantum computers over two days while maintaining low mean TVD versus other methods.}
    \label{fig:real_machine}
    \vspace{-4mm}
\end{figure}

\noindent\textbf{Real-Machine Evaluation on Superconducting-Qubit Hardware:} To evaluate the effectiveness of \sol{} in realistic deployment scenarios, we benchmarked its performance on actual superconducting-qubit hardware. Specifically, we executed the VAR algorithm multiple times across five of IBM’s latest 127-qubit quantum processors—namely, \textit{ibm\_sherbrooke}, \textit{ibm\_kyiv}, \textit{ibm\_brisbane}, \textit{ibm\_brussels}, and \textit{ibm\_strasbourg}. These runs were distributed over a two-day period to capture the temporal variability inherent to cloud-based quantum hardware. Figure~\ref{fig:real_machine} summarizes the results. Consistent with trends observed in our simulation-based experiments, \sol{} demonstrates strong robustness under real hardware conditions. Across all machines and execution windows, \sol{} achieves the lowest TVD standard deviation of $5.9\%$, significantly outperforming other techniques such as OptiMap ($8.8\%$), RandMap ($17.8\%$), and EqualDist ($10.3\%$). This reduced variance indicates that \sol{} delivers more stable output distributions across runs. \sol{} also maintains a favorable mean TVD of $19.6\%$, which is on par with the more optimized OptiMap method ($18.4\%$), while substantially outperforming RandMap ($44.0\%$) and EqualDist ($42.5\%$). These results indicate that \sol{} not only offers consistent behavior over time but also produces high-fidelity outputs, making it a compelling candidate for practical deployment in noisy hardware.

\vspace{2mm}

\begin{figure}[t]
    \centering
    \includegraphics[width=0.99\columnwidth]{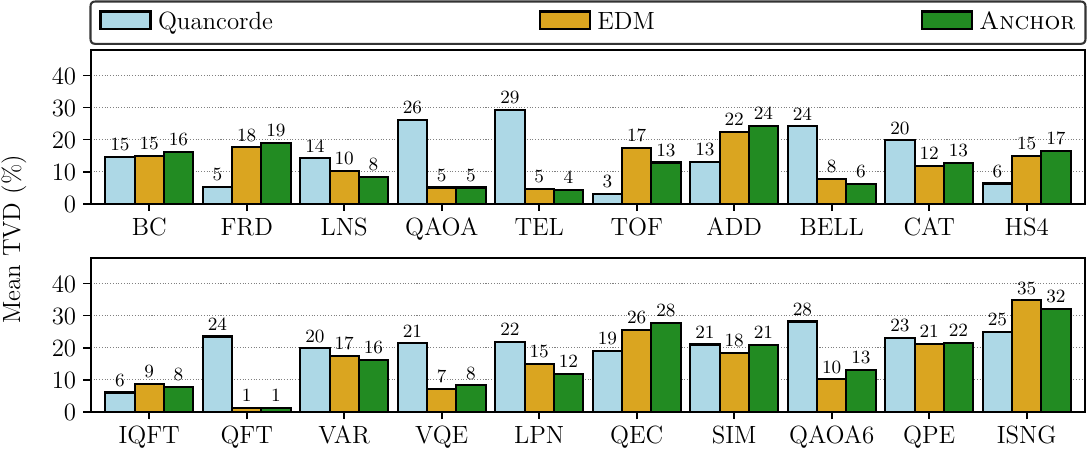}
    \vspace{1mm}
    \hrule
    \vspace{-3.5mm}
    \caption{\sol{} achieves a lower or comparable mean TVD on average as compared to state-of-the-art techniques to reduce the mean TVD.}
    \label{fig:mean_tvd_edm_quancorde}
    \vspace{-4mm}
\end{figure}

\begin{figure}[t]
    \centering
    \includegraphics[width=0.99\columnwidth]{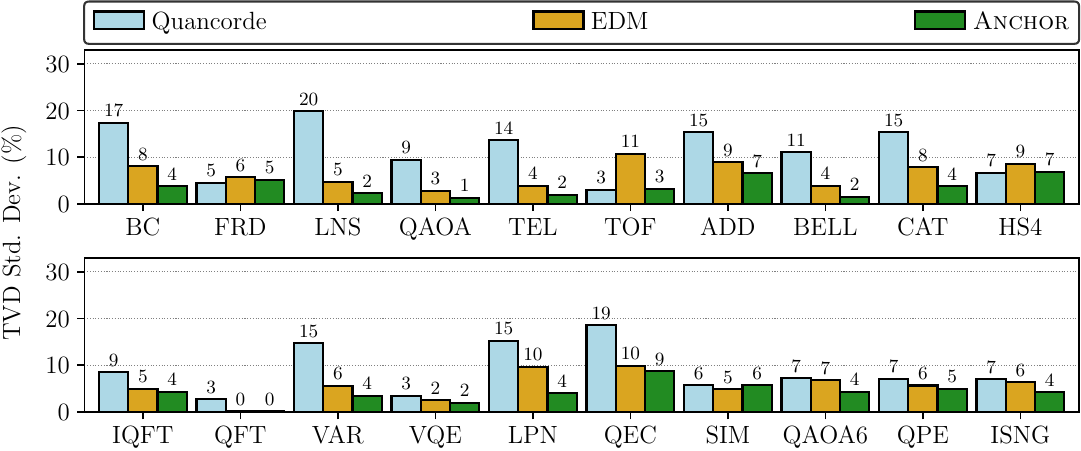}
    \vspace{1mm}
    \hrule
    \vspace{-3.5mm}
    \caption{\sol{} achieves a lower TVD standard deviation on average as compared to state-of-the-art techniques to reduce the TVD standard deviation.}
    \label{fig:std_tvd_edm_quancorde}
    \vspace{-4mm}
\end{figure}

\noindent\textbf{Performance Comparison with Variability-agnostic State-of-the-art Techniques}. Lastly, we compare \sol{} to state-of-the-art techniques that are focused on improving the average fidelity, but not directly the variability in the fidelity across multiple runs. We compare against two techniques: (1) Quancorde~\cite{ravi2022quancorde} and (2) EDM~\cite{tannu2019ensemble}. Both of these are ensemble-based techniques, meaning they run multiple circuit maps across different devices for each algorithm and therefore, might be able to reduce variability even if they do not directly intend to. Quancorde uses an ensemble of classically simulable Clifford versions of the original circuit to get a noise estimate of the original circuit. EDM uses an ensemble of diverse maps to run a circuit to attempt to cancel out uncorrelated noise across different maps. Our results on mean TVD are presented in Fig.~\ref{fig:mean_tvd_edm_quancorde}, and TVD standard deviation are presented in Fig.~\ref{fig:std_tvd_edm_quancorde}. The results show that Quancode has a much higher mean TVD on average, while EDM and \sol{} have comparable mean TVDs. On the other hand, \sol{} has the lowest TVD standard deviation for all the algorithms as compared to Quancorde and EDM. This supports \sol{}'s campaign of directly targeting the variability in fidelity instead of only relying on techniques that attempt to improve the mean fidelity.

%% file: sections/related_work.tex
\section{Related Work}
\label{sec:related_work}

Prior work relevant to \sol{} can be classified into three broad categories:

\vspace{2mm}

\noindent\textbf{(1) Optimal Quantum Circuit Mapping Techniques:} Much of the work geared toward improving quantum circuits' performance when run on noisy quantum computers has focused on reducing the estimated output error. This is done by optimally mapping and routing the circuit for the measured qubit errors during a calibration cycle~\cite{tannu2019ensemble,tannu2019not,wille2019mapping,zulehner2019compiling,patel2020ureqa,patel2020veritas,gokhale2020optimized,li2019tackling}. As an instance, one of the first works in this area by Tannu et al.~\cite{tannu2019not} highlighted the fact that different qubits have different error rates, and it would be beneficial to map a circuit carefully to reduce the overall output error. Compiler optimizations~\cite{10.1145/3669940.3707240} reduce circuit gate counts but are not designed to address quantum hardware's temporal and spatial variations. A few works have also proposed mapping a circuit to multiple computer regions and averaging their outputs~\cite{patel2020veritas,tannu2019ensemble}. However, this is proposed as a methodology to reduce the overall error, not the variability across different runs. Nonetheless, as we showed in our evaluation (Sec.~\ref{sec:evaluation}), a technique similar to this approach (EqualDist) does not perform well compared to \sol{}. 


\vspace{2mm}

\noindent\textbf{(2) Quantum Error Estimation and Prediction:} The estimation of the output error of a quantum circuit is widely done using a metric called Estimated Success Probability (ESP)~\cite{patel2023graphine,li2022optimal,brandhofer2023optimal,tannu2019ensemble,xie2021mitigating,patel2022geyser}. It is calculated by taking a product of the success rates of individual gates in the quantum circuit and the decoherence times of the qubits. However, in our work, we found this metric to be a poor estimator of the TVD for superconducting systems as it does not take into account complex noise interactions. Due to this, prior work by Patel et al.~\cite{patel2021qraft} has also attempted to develop a machine-learning-based predictor for the output error. However, this work assumes the circuit was already run and uses that output to correct the error for individual states, which is not suitable for the purpose of \sol{} as it needs to predict the TVD without running the circuit at all.

\vspace{2mm}

\noindent\textbf{(3) Quantum Computer Characterization and Benchmarking:} Prior research has also explored designing different benchmarking algorithms for quantum computers~\cite{li2023qasmbench,tomesh2022supermarq,bassman2021arqtic} and characterizing the behavior of quantum hardware using studies that span multiple algorithms, computers, and days~\cite{ranjan2023experimental,patel2020experimental,ravi2021quantum}. For instance, Ravi et al.~\cite{ravi2021quantum} demonstrate how the performance of a quantum algorithm depends on the chosen quantum hardware and the timing of the run. Overall, these studies have also shown a high degree of variability in the output error of quantum circuits when run on quantum computers, which in part inspired the conception of \sol{}. However, unlike \sol{}, these works do not propose a solution to the variability challenge.

\vspace{-0.5mm}

%% file: sections/conclusion.tex
\section{Conclusion}
\label{sec:conclusion}

In this paper, we have presented \sol{}, a novel technique designed to address the significant challenge of performance variability on NISQ computers. Our approach leverages linear programming and a performance predictor to reduce the variability in output fidelity, providing more consistent performance for the same quantum circuit across different quantum computers and over different time periods. This marks a departure from current methods that primarily focus on error reduction without addressing the inherent variability in quantum circuit outputs. Through rigorous evaluation, \sol{} has demonstrated an average reduction in performance variability by 73\% compared to state-of-the-art implementations. Our work not only contributes a valuable tool for mitigating performance variability but also sets the stage for future research in this critical area. \sol{} opens up new possibilities for the advancement of quantum computing technologies, moving us closer to realizing the full potential of quantum acceleration in practical applications. \rev{\sol{}'s framework, codebase, and experimental data are open-sourced for reproducibility and research community adoption at \textit{\url{https://github.com/positivetechnologylab/Anchor}}.}

\section*{\rev{Acknowledgement}}

\rev{We would like to thank the anonymous reviewers and our shepherd, Professor Thirupathaiah Vasantam, for their valuable and insightful feedback that has helped improve this work. This work was supported by Rice University, the Rice University George R. Brown School of Engineering and Computing, and the Rice University Department of Computer Science. This work was supported by the DOE Quantum Testbed Finder Award DE-SC0024301, the Ken Kennedy Institute, and Rice Quantum Initiative, which is part of the Smalley-Curl Institute. We acknowledge the use of IBM Quantum services for this work. The views expressed are those of the authors, and do not reflect the official policy or position of IBM or the IBM Quantum team.}



%% file: sections/appendix.tex
\section{\rev{Worked Example: From Maps to Predicted TVD to LP to Shot Allocation}}
\label{sec:example}

\rev{We illustrate the end-to-end \sol{} flow on a toy 2-qubit circuit with two candidate maps per computer. 
\sol{} first generates diverse circuit maps (\textsc{OptiMap} and \textsc{RandMap}) for each computer, predicts the TVD for each (via the TVD predictor), and then solves a linear program (LP) to pick shot fractions per map that both (i) minimize overall expected TVD and (ii) equalize the expected TVD across computers (up to tolerance~$\epsilon$). Finally, only the map fractions for the scheduled computer are executed; other computers only serve to shape the equalization target.}

\vspace{2mm}

\noindent\rev{\textbf{Setup:} Suppose a user's job is scheduled on \emph{Computer~1} (the only place we will actually run shots). 
For clarity, we consider two computers ($k{=}2$) and two maps per computer ($m{=}2$), named $A$ and $B$. $A$ is generated using \textsc{OptiMap} and $B$ is generated using \textsc{RandMap}.
Let $\mathrm{TVD}_{iA}$ and $\mathrm{TVD}_{iB}$ denote the predictor’s expected TVD for maps $A$ and $B$ on \emph{Computer~$i$}. 
For this example, the TVD predictor produces the following table (numbers in $[0,1]$):
\[
\begin{array}{c|cc}
\text{Computer} & \text{Map }A & \text{Map }B \\ \hline
1 & 0.12 & 0.22 \\
2 & 0.18 & 0.14 \\
\end{array}
\]
Intuitively, on \emph{Computer~1} map $A$ is better, while on \emph{Computer~2} map $B$ is better.}

\vspace{2mm}

\noindent\rev{\textbf{LP Variables and Constraints:} Let $f_{ij}\in[0,1]$ be the fraction of shots assigned to map $j\!\in\!\{A,B\}$ on computer $i\!\in\!\{1,2\}$. 
Define the expected TVD on computer $i$ as 
\[
\mathrm{TVD}_i \;=\; f_{iA}\,\mathrm{TVD}_{iA} + f_{iB}\,\mathrm{TVD}_{iB},
\quad \text{with } f_{iA}+f_{iB}=1 \text{ for each } i.
\]
\sol{}’s LP (i) \emph{minimizes} a linear objective $c^\top x$ built from the predicted TVDs, and (ii) enforces \emph{equalization} across computers via (soft) pairwise bounds:
\[
\frac{1}{1+\epsilon}\,\mathrm{TVD}_i \;\le\; \mathrm{TVD}_j \;\le\; (1+\epsilon)\,\mathrm{TVD}_i \quad \text{for all pairs } (i,j).
\]
Concretely, the LP is:
\[
\min_{x} \; c^\top x \quad 
\text{s.t. } A_{\mathrm{eq}}x=b_{\mathrm{eq}},\; A_{\mathrm{ub}}x \le b_{\mathrm{ub}},\; 0\le x\le 1,
\]
where $x=[f_{1A},f_{1B},f_{2A},f_{2B}]^\top$ and $c=[\mathrm{TVD}_{1A},\mathrm{TVD}_{1B},\mathrm{TVD}_{2A},\mathrm{TVD}_{2B}]^\top = [0.12, 0.22, 0.18, 0.14]^\top$ (See Sec..~\ref{sec:design} for how this form is derived). We thus get the following:
\[
A_{\mathrm{eq}}
=
\begin{bmatrix}
1 & 1 & 0 & 0 \\[4pt]
0 & 0 & 1 & 1
\end{bmatrix},
\qquad
b_{\mathrm{eq}}
=
\begin{bmatrix}
1 \\[4pt]
1
\end{bmatrix},
\]
\[
A_{\mathrm{ub}}
=
\begin{bmatrix}
-(1+\epsilon)\cdot 0.12 & -(1+\epsilon)\cdot 0.22 & 0.18 & 0.14 \\[6pt]
\frac{0.12}{1+\epsilon} & \frac{0.22}{1+\epsilon} & -0.18 & -0.14 \\[6pt]
0.12 & 0.22 & -(1+\epsilon)\cdot 0.18 & -(1+\epsilon)\cdot 0.14 \\[6pt]
-0.12 & -0.22 & \frac{0.18}{1+\epsilon} & \frac{0.14}{1+\epsilon}
\end{bmatrix},
\qquad
b_{\mathrm{ub}}
=
\begin{bmatrix}
0 \\[4pt] 0 \\[4pt] 0 \\[4pt] 0
\end{bmatrix}.
\]
Combined, we get:
\[
A =
\begin{bmatrix}
1 & 1 & 0 & 0 \\
0 & 0 & 1 & 1 \\ \hline
-(1+\epsilon)\cdot 0.12 & -(1+\epsilon)\cdot 0.22 & 0.18 & 0.14 \\
\frac{0.12}{1+\epsilon} & \frac{0.22}{1+\epsilon} & -0.18 & -0.14 \\
0.12 & 0.22 & -(1+\epsilon)\cdot 0.18 & -(1+\epsilon)\cdot 0.14 \\
-0.12 & -0.22 & \frac{0.18}{1+\epsilon} & \frac{0.14}{1+\epsilon}
\end{bmatrix},
\qquad
b =
\begin{bmatrix}
1 \\ 1 \\\hline 0 \\ 0 \\ 0 \\ 0
\end{bmatrix}.
\]}

\vspace{2mm}

\noindent\rev{\textbf{Solving the Toy Instance:} Take $\epsilon{=}0$ for simplicity. Then, we have to solve for:
\[
\min_{x} \;
\begin{bmatrix}
0.12 & 0.22 & 0.18 & 0.14
\end{bmatrix}
\begin{bmatrix}
f_{1A} \\ f_{1B} \\ f_{2A} \\ f_{2B}
\end{bmatrix}
\quad s.t. \quad
\begin{bmatrix}
1 & 1 & 0 & 0 \\
0 & 0 & 1 & 1 \\
-0.12 & -0.22 & 0.18 & 0.14 \\
0.12 & 0.22 & -0.18 & -0.14 \\
0.12 & 0.22 & -0.18 & -0.14 \\
-0.12 & -0.22 & 0.18 & 0.14
\end{bmatrix}
\begin{bmatrix}
f_{1A} \\ f_{1B} \\ f_{2A} \\ f_{2B}
\end{bmatrix}
=
\begin{bmatrix}
1 \\ 1 \\ 0 \\ 0 \\ 0 \\ 0
\end{bmatrix}.
\]
Equalization requires $\mathrm{TVD}_1 = \mathrm{TVD}_2$ at optimum. 
On \emph{Computer~2}, the best (lowest) TVD is achieved by $f_{2A}{=}0,\;f_{2B}{=}1$, yielding $\mathrm{TVD}_2 = 0.14$. 
To match this on \emph{Computer~1}:
\[
f_{1A}\cdot 0.12 + (1-f_{1A})\cdot 0.22 \;=\; 0.14 
\;\;\Rightarrow\;\;
-0.10 f_{1A} + 0.22 = 0.14 
\;\;\Rightarrow\;\;
f_{1A}=0.80,\;\; f_{1B}=0.20.
\]
Thus, the LP returns (one optimal solution):
\[
(f_{1A},f_{1B};\; f_{2A},f_{2B}) \;=\; (0.80,\,0.20;\; 0,\,1.00),
\quad \text{with } \mathrm{TVD}_1=\mathrm{TVD}_2=0.14.
\]}

\vspace{0mm}

\noindent\rev{\textbf{Shot Allocation:} \sol{} executes \emph{only} the scheduled computer’s fractions. 
If the job asked for $32{,}000$ shots, we run them on \emph{Computer~1} as:
\[
\text{Map }A:\; 0.80\times 32{,}000 = 25{,}600 \text{ shots}, 
\qquad
\text{Map }B:\; 0.20\times 32{,}000 = 6{,}400 \text{ shots}.
\]
No shots are run on \emph{Computer~2}; it is used to set the equalization target. 
This achieves both low expected TVD and cross-computer equalization, reducing spatial/temporal variability while keeping execution localized to the scheduled backend.}

\vspace{2mm}

\noindent\rev{\textbf{Notes:} (i) With $\epsilon{>}0$, the LP permits small controlled deviations from perfect equality if that lowers the objective; (ii) with more maps/computers, the LP naturally generalizes; (iii) if a single map dominates on all computers, the solution reduces to that map; and (iv) the numeric values here are illustrative but follow the same path \sol{} uses at scale.}